\documentclass[aps,prl,reprint, superscriptaddress,preprintnumbers, amsmath,amssymb,nofootinbib]{revtex4-1}
\usepackage{graphicx}
\usepackage[dvipsnames]{xcolor}
\usepackage{hyperref}
\usepackage{xspace}
\usepackage{ifdraft}
\usepackage{epstopdf}
\usepackage{slashed}

\definecolor{jaxoblue}{HTML}{0086FF}

\newcommand{\cC}{g} 

\newcommand{\xid}{X identity}
\newcommand{\hid}{H identity}

\newcommand{\nkmc}[1][k]{N$^{#1}$MC\xspace}

\newcommand{\locp}{\ensuremath{\mathcal{P}_{\gamma^{(k)}}}}
\newcommand{\cgamma}{\ensuremath{\mathcal{C}_{\gamma^{(k)}}}}
\newcommand{\ratr}{\ensuremath{\mathcal{R}_{\gamma^{(k)}}}}

\newcommand{\arrdiag}[2]{\begin{minipage}{#1}\includegraphics[width=\textwidth]{#2}\end{minipage}}

\DeclareMathOperator{\cut}{\mathcal{C}}

\def\spa#1.#2{\left\langle#1\,#2\right\rangle}
\def\spb#1.#2{\left[#1\,#2\right]}
\def\spash#1.#2{\spa{\smash{#1}}.{\smash{#2}}}
\def\spbsh#1.#2{\spb{\smash{#1}}.{\smash{#2}}}
\def\sand#1.#2.#3{%
\left\langle\smash{#1}{\vphantom1}^{-}\right|{#2}%
\left|\smash{#3}{\vphantom1}^{-}\right\rangle}
\def\sandpp#1.#2.#3{%
\left\langle\smash{#1}{\vphantom1}^{+}\right|{#2}%
\left|\smash{#3}{\vphantom1}^{+}\right\rangle}
\def\sandpm#1.#2.#3{%
\left\langle\smash{#1}{\vphantom1}^{+}\right|{#2}%
\left|\smash{#3}{\vphantom1}^{-}\right\rangle}
\def\sandmp#1.#2.#3{%
\left\langle\smash{#1}{\vphantom1}^{-}\right|{#2}%
\left|\smash{#3}{\vphantom1}^{+}\right\rangle}

\def\Tr{\, {\rm Tr}}

\def\eqn#1{eq.~(\ref{#1})}

\def\be{\begin{equation}}
\def\ee{\end{equation}}
\def\bea{\begin{eqnarray}}
\def\eea{\end{eqnarray}}
\def\ba{\begin{eqnarray}}
\def\ea{\end{eqnarray}}

\newcommand{\afour}{\ensuremath{A_4^{\text{tree}}}}

\begin{document}

\preprint{
UUITP-62/21
}

\author{John Joseph M. Carrasco}
\affiliation{Department of Physics and Astronomy, Northwestern
  University, Evanston, Illinois 60208, USA}
\affiliation{Institut de Physique Th\'{e}orique, Universite Paris Saclay, CEA, CNRS, F-91191 Gif-sur-Yvette, France}
\author{Alex Edison}
\affiliation{
Department of Physics and Astronomy, Uppsala University, Box 516,  75120 Uppsala, Sweden}
\author{Henrik Johansson}
\affiliation{
Department of Physics and Astronomy, Uppsala University, Box 516,  75120 Uppsala, Sweden}
\vspace{0.9cm}
\affiliation{Nordita, Stockholm University and KTH Royal Institute of Technology,  Hannes Alfv\'{e}ns  v\"{a}g 12, 10691 Stockholm, Sweden}

\title{Maximal Super-Yang-Mills at Six Loops via Novel Integrand Bootstrap}

\begin{abstract}
  We construct the complete (planar and non-planar) integrand for the
  six-loop four-point amplitude in maximal $D\le10$ super-Yang-Mills.
  This construction employs new advances that combat the proliferation
  of diagram contributions and state sums when evaluating multi-loop $D$-dimensional
  unitarity cuts.  Concretely, we introduce two graph-based
  approaches to evaluating
  generalized unitarity cuts in $D$ dimensions: 1) recursively from
  lower-loop cuts, or 2) directly from known higher-loop planar cuts.
  Neither method relies on explicit state sums or any sewing of
  tree-level amplitudes.  These methods are based on identities that we expect to hold for a broad family of theories, including  QCD and Einstein gravity.
\end{abstract}

\maketitle

\section{Introduction}

Gauge theory and general relativity are centerpieces in the
theoretical framework that describes modern physics. Tremendous theoretical and experimental effort has gone into improving our understanding of non-abelian
gauge theory. In comparison, much less is known about gravity, at the quantum level. A primary reason for this disparity is the lack of experiments that probe
relevant energy scales. A secondary reason is the technical challenge of working with the highly non-linear behavior of gravity.
General relativity, as a quantum theory, is famously known to be both non-renormalizable and ultraviolet (UV) divergent~\cite{Goroff:1985th}, and the question of its UV-completion is an all-important open problem.

A longstanding well-behaved theory, that extends general relativity
with a finite number of fields without higher-derivative interactions, is the maximally supersymmetric
${\cal N}=8$ supergravity, discovered by Cremmer, Julia and
Scherk~\cite{Cremmer:1978km}. It is expected to be UV finite in four dimensions up to at
least six loops, with the first counterterm compatible with known
symmetries appearing at seven loops~\cite{Kallosh:2009db,Bossard:2009sy,Howe:1980th,Kallosh:1980fi,Berkovits:2006vc,Green:2006yu,Green:2006gt,Marcus:1984ei,Berkovits:2009aw,Vanhove:2010nf,Beisert:2010jx,Bossard:2010bd,Bjornsson:2010wu,Bossard:2011tq}. In the
absence of a proof of finiteness, the only available means
of showing the presence or absence of this divergence is by an
explicit seven-loop calculation.

Progress in maximal supergravity hinges directly on
developments in ${\cal N}=4$ super-Yang-Mills (SYM) theory, first written down by
Brink, Schwarz and Scherk~\cite{Brink:1976bc}. The theories are
related by the double copy at the
classical~\cite{Kawai:1985xq,Bern:2008qj} and quantum
level~\cite{Bern:2010ue},
\begin{equation}
\big({\cal N}=8~\text{supergravity}\big)~ = ~\big({\cal N}=4~\text{SYM}\big)^2\,.
\end{equation}
Consider the asymptotic spectra: ${\cal N}=4$
SYM has 8 bosonic plus 8 fermionic states, and the
${\cal N}=8$ supergravity states are the square of these
$(16)^2=256$. The double copy, in combination with the
unitarity method~\cite{Bern:1994zx,Bern:1994cg}, has been successfully used to determine
the ${\cal N}=8$ supergravity amplitudes through five
loops~\cite{Bern:2007hh,Bern:2008pv, Bern:2009kd,Bern:2012uf,
  Bern:2017yxu,Bern:2017ucb,Bern:2018jmv}.  However, the
progress relied, in part, on the powerful constraints implied by color-kinematics duality~\cite{Bern:2008qj,Bern:2010ue}, which obtains the complete integrand from only a small subset of all diagrams and cuts~\cite{Bern:2010ue,Carrasco:2011hw,Bern:2012uf}.

A generalized double copy~\cite{Bern:2017ucb,Bern:2017yxu}  was introduced to ameliorate challenges with manifesting color-kinematics duality at five loops, and while it was sufficiently powerful to complete the calculation~\cite{Bern:2018jmv}, new methods and refinements are clearly needed to tackle the six- and seven-loop calculations. 
At high loop orders, apparently mundane tasks can become impenetrable walls, such as evaluating unitarity cuts. In ${\cal N}=4$
SYM, one faces the problem of exponentially growing state sums with $(16)^L$ terms at $L$ loops.  Alternatively, state sums can be handled covariantly in ten-dimensional notation, but the fermionic Dirac traces resist evaluation; for $L=6$, one expects traces of up to 28 gamma matrices. Additionally, the approach obscures the enormous cancellations due to supersymmetry.

In this Letter, we introduce two new and complementary bootstrapping methods capable of overcoming obstacles encountered in ascending towards seven loops. The methods circumvent the need for laborious state sums when evaluating multi-loop unitarity cuts, and they manifest symmetries (supersymmetry, $D$-dimensional Poincar\'{e} symmetry) that may otherwise rely on non-trivial cancellations. The general idea is to relate unknown cuts to known simpler cuts, and recursively bootstrap the integrand from tree level. The two approaches indirectly exploit that gauge-theory amplitudes can be stratified in two complementary ways: the loop and 't Hooft expansions~\cite{tHooft:1973alw}.

The first relationship, which we call the \xid,  describes the degeneration of a planar four-point tree amplitude into
two disconnected non-planar lines.  Embedded within a larger cut, it determines non-planar cuts from limits of
\emph{higher-loop} \emph{planar} cuts.  Planar cuts tend to be more
straightforward~\cite{Bern:2006ew,Bern:2007ct,Arkani-Hamed:2010zjl,Bern:2012di,Bourjaily:2016evz}, and can be highly constrained; e.g.~in ${\cal N}=4$
SYM, by dual-conformal~\cite{Drummond:2006rz,Drummond:2008vq,Brandhuber:2008pf}, Yangian~\cite{Drummond:2009fd} symmetry, or Grassmannian and Amplituhedron geometry~\cite{Arkani-Hamed:2012zlh,Arkani-Hamed:2013jha,Arkani-Hamed:2014dca,Arkani-Hamed:2018rsk,Langer:2019iuo}.  

The second relationship, called the \hid,  similarly relates a degeneration limit of the factorized four-point amplitude to two disconnected planar lines.  Applied to a multi-loop cut, the H-identity allows for the bootstrap of cuts in terms of known \emph{lower-loop} cuts, while maintaining the planarity properties (crossing count) of graphs.

As a demonstration of the usefulness of the new bootstrapping techniques, we apply them together with the method of maximal cuts
\cite{Bern:2007ct,Bern:2008pv,Bern:2010tq,Bern:2012uf,Bern:2018jmv}, to construct the complete six-loop four-point $D$-dimensional integrand in ${\cal N}=4$ SYM. This constitutes the first non-planar computation at six loops in ${\cal N}=4$ SYM, thus extending the planar six-loops results~\cite{Bourjaily:2011hi,Eden:2012tu,Bern:2012di}.  We anticipate the new methods introduced here to be highly useful in many other theories relevant to modern physics, including QCD and Einstein gravity. 

\section{Deriving  the X-ing and H-ing moves}
\noindent
{\bf \xid:} Take $\langle 1^a,2^b | S | 3^c ,4^d \rangle$ to be a tree-level $S$-matrix element in some Yang-Mills gauge theory, where the states
are distinguished by their momenta $p_1+p_2=p_3+p_4$ as well as
Lorentz (little group) representations, which we indicate using generalized indices
$a,b,c,d$. The particles also transform in representations of the
gauge group, but the details are not needed for this argument.  We only
need to assume that the $S$-matrix elements can be expanded in terms of
partial amplitudes that have a notion of planarity, analogous to purely-adjoint theories.

\begin{figure}[ht]
  \arrdiag{0.10\textwidth}{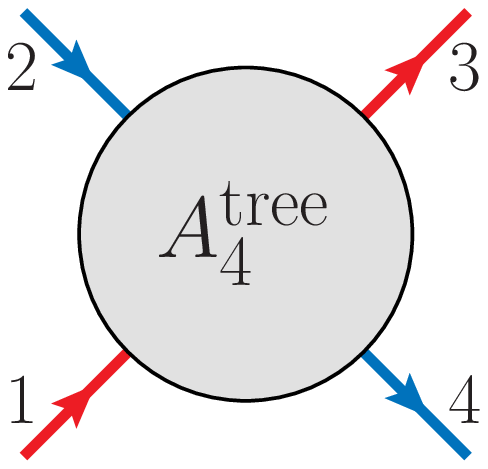} = \arrdiag{0.10\textwidth}{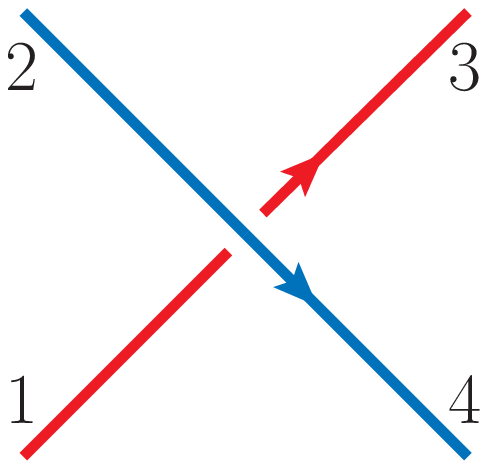}
  \caption{\label{fig1}Diagrammatic interpretation of the \xid.}
\end{figure}

Thus for our purposes, it is convenient to strip off most factors (color, coupling, momentum-conserving delta functions, phases) and
consider a planar tree-level amplitude $A(1^a,2^b | 3^c,4^d)$. We
note that this amplitude, which is a dimensionless function that is covariant in the generalized indices, must have
a simple behavior in the kinematic limit $p_3\rightarrow p_1$,
which has but one scale, $s_{12}= (p_1+p_2)^2$.  In a suitably
chosen normalization, we expect it to either evaluate to 0 or 1.  It is not difficult to see in explicit examples
that the correct answer is
\begin{equation} \label{X-identity}
\afour(1^a,2^b | 3^c,4^d)\big|_{p_3\rightarrow p_1} = \delta^{ac} \delta^{bd}\,,
\end{equation}
where $ \delta^{ab}$ simply identifies the states
on diagonally opposite legs. We will call this equation the
\xid, and it is graphically represented in Fig.~\ref{fig1}.

Let us illuminate \eqn{X-identity} by considering $D$-dimensional pure SYM theory, described by the adjoint fields $\{A^\mu, \psi \}$
and the Lagrangian
$\Tr[-\frac{1}{4}(F^{\mu\nu})^2+ \frac{i}{2} \bar \psi \slashed{D}
\psi]$. In the given limit, the non-vanishing partial amplitudes are
\begin{align}
  A_{\text{SYM}}^{\text{tree}}(1_A,2_A|3_A,4_A)\big|_{p_3 \to p_1} &= (\varepsilon_1 \cdot \varepsilon_3)(\varepsilon_2 \cdot \varepsilon_4)\,, \nonumber \\
  A_{\text{SYM}}^{\text{tree}}(1_\psi,2_A|3_\psi,4_A)\big|_{p_3 \to p_1} &= (\bar \chi_1 \chi_3)(\varepsilon_2 \cdot \varepsilon_4)\,,\nonumber \\
  A_{\text{SYM}}^{\text{tree}}(1_\psi,2_\psi |3_\psi,4_\psi)\big|_{p_3 \to p_1} &= (\bar \chi_1 \chi_3) (\bar \chi_2 \chi_4)\,,
\end{align}
where $\varepsilon^\mu$ and $\chi$ are dimensionless polarizations of
respective states (for convenience, the fermion wavefunctions are
normalized as $\bar{\chi}\chi = 1$ here).  All other amplitudes either
vanish, or are trivially related to the above ones by permutations of
legs, in accordance with \eqn{X-identity}. The simple result
for the four-fermion amplitude relies on the same Fiertz identity that
is responsible for supersymmetry in the theories under consideration.
In our case, we have $D=3,4,6,10$ as valid dimensions where the
\xid{} holds in term of the SYM fields $\{A^\mu, \psi
\}$. Dimensional compactifications of these theories will also obey
\eqn{X-identity}, which will introduce scalars and possibly mass
terms in the SYM spectrum. In $D=4$, it follows that the
${\cal N}=0,1,2,4$ SYM theories obey the \xid{} (with possible mass terms compatible with spontaneous symmetry breaking or dimensional compactification). 

The \xid{} (\ref{X-identity}), simple as it is, has profound consequences. It is sufficient to take a kinematic limit in order
for the planar tree amplitude to collapse into a non-planar identity
operator. The fact that this can hold for arbitrary states in large families of gauge theories presages powerful promise for loop calculation.

Assume that we know the planar $L$-loop $n$-point
amplitude integrand in a gauge theory. Using the \xid, we can
directly compute non-planar $n$-point cuts with \emph{more} edge
crossings, but \emph{lower} loops via taking appropriate kinematic
limits on planar cuts with quartic vertices.  Using $\cC$ to refer to the
number of non-planar crossings, we can write a higher-loop cut for a graph $\gamma_{\rm X}$ that contains an isolated quartic interaction as
\begin{equation}
  \cut^{\cC,L}_{\gamma_{\rm X}}  \equiv \widehat{\cut}{}^{\cC}(\ell_4^d,\ell_3^c|\ell_2^b,\ell_1^a) \afour(\ell_1^a,\ell_2^b|\ell_3^c, \ell_4^d)
\end{equation}
where the four-point tree is factored out, repeated indices are summed over, and external states are not shown.  Applying
the \xid{} (\ref{X-identity}) on this cut yields
\begin{align}
  \lim_{\ell_3 \to \ell_1} \cut^{\cC,L}_{\gamma_{\rm X}} \! &=  \widehat{\cut}{}^{\cC}(\ell_4^d,\ell_3^c|\ell_2^b,\ell_1^a) \afour(\ell_1^a,\ell_2^b|\ell_3^c, \ell_4^d)\Big|_{\ell_3 \to \ell_1} \notag \\
  &= \widehat{\cut}{}^{\cC}(\ell_2^b,\ell_1^a|\ell_2^b,\ell_1^a) \equiv \cut^{\cC+1,L-1}\,,
    \label{eq:x-id}
\end{align}
where in the final step a generic cut is obtained, with two on-shell loop legs sewn in a non-planar fashion.  This final sewing lowers the loop order at the cost of introducing a new crossing. 
(Equivalent relations apply if some of the legs of the four-point sub-amplitude are external, but the details of the crossing and loop counting may differ.)  To reach an $L$-loop cut with $\cC$ edge crossings, we simply start with an $(L+\cC)$-loop cut with $\cC$ four-point sub-amplitudes and apply the X-identity to each of those sub-amplitudes. In this way, any physical cut in a gauge theory can be straightforwardly computed from the planar cuts, assuming they are known. 

\vskip0.1cm
\noindent
{\bf \hid:} Since the loop-level \xid{} requires knowledge of higher loop (planar) integrands, there is a non-trivial threshold to pass before using it. However, we now consider a similar identity, derivable from the \xid, that admits a bootstrap of integrands from lower loops without necessarily adding non-planar crossings. 

\begin{figure}[ht]
  $s_{23}\times$\!\arrdiag{0.10\textwidth}{figs/x-rule-tree}\!\! $\xrightarrow{p_1 \to p_3}$ \!\!\arrdiag{0.10\textwidth}{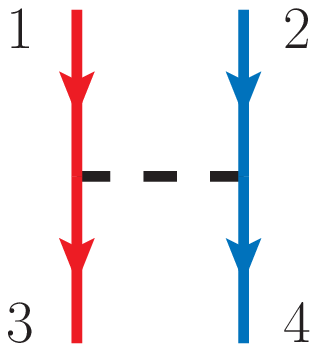}\!\! $= s_{12}\times$\!\arrdiag{0.10\textwidth}{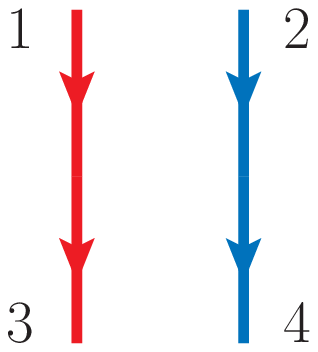}
  \caption{\label{fig2}Diagrammatic interpretation of the \hid{} (right), and its connection to the \xid{} (left).
    The dashed line represents an on-shell state with vanishing momentum.}
\end{figure}

At tree-level the \hid{} takes the form 
\begin{equation} \label{H-identity} 
  \lim_{k \to 0} A_{3}^{\text{tree}}(1^a|3^c,k^e) A_3^{\text{tree}}(2^b,k^e|4^d) \!=\! s_{12}\delta^{ac}\delta^{bd}\,,
\end{equation}
where, as before, repeated indices are summed over, and momentum is conserved for each three-point amplitude ($p_1-p_3=k=p_4-p_2$). The diagrammatic interpretation of the \hid{} is given in the right-side pannel of
Fig.~\ref{fig2}. 
The left-side pannel of the same figure indicates how it is derived from the  \xid. It is obtained from applying
the \xid{} to a BCJ relation \cite{Bern:2008qj}
\begin{equation} \label{BCJrel}
  \lim_{p_3 \to p_1} \Big\{s_{13} \afour(1,2,4,3)\! =\! s_{23}\afour(1,2,3,4) \Big\} \,.
\end{equation}
Here the left hand side evaluates to the residue on the $s_{13}=(p_1+p_3)^2$ pole, yielding the factorized three-point amplitudes, and using \eqn{X-identity} the right-hand-side amplitude
becomes the product of delta functions.

The \hid{} is straightforward to apply to any cut that has a connected pair of three-point sub-amplitudes
\begin{equation}
  \mathcal{C}_{\gamma_H}^{L} \! \equiv \widehat{\cut}(\ell_4^d,\ell_2^b|\ell_3^c,\ell_1^a) A_3^{\text{tree}}(\ell_1^a|\ell_3^c,\ell_5^e)A_3^{\text{tree}}(\ell_2^b, \ell_5^e| \ell_4^d)
\end{equation}
yielding a constraint equation satisfied by the cut
\begin{align}
   \lim_{\ell_5 \to 0} \mathcal{C}_{\gamma_{\rm H}}^{L} &= 2 \ell_{1} {\cdot} \ell_{2}\, \mathcal{C}^{L{-}1}_{\gamma_{\rm H} \setminus \ell_5} \,,
\label{eq:h-id}
\end{align}
where $\mathcal{C}^{L{-}1}_{\gamma_{\rm H} \setminus \ell_5}$ is a lower-loop cut that is assumed to be known. 
While the \hid{} does not calculate the cut $\mathcal{C}_{\gamma_{\rm H}}^L$ for general momenta, the full set of such conditions is often powerful enough for a complete determination, as will be discussed below for the six-loop $\mathcal{N}=4$ SYM calculation. One might also worry that the set of cuts with two connected tree-point sub-amplitudes is too special.  However, in the maximal cut method most of the needed cuts are of this form, and for $\mathcal{N}=4$ SYM these should be sufficient to determine the full amplitude. 

Note that while the \hid{} is reminiscent of the rung rule of planar $\mathcal{N}=4$ SYM~\cite{Bern:1997nh,Bern:1998ug}, here we stress that the \hid{} is not a heuristic rule, and it applies both to non-planar amplitudes, and a multitude of theories (including those theories mentioned above for the \xid). Its generality can be traced back to the universality of soft factorization~\cite{Weinberg:1964ew,Weinberg:1965nx}, which is a general feature of gauge~\cite{Low:1954kd, Gell-Mann:1954wra} and gravity theories~\cite{Gross:1968in} (see also~\cite{Laenen:2008gt,White:2011yy}). Indeed, the \hid{} has a direct extension to a multitude of gravitational theories, simply by replacing $s_{12}\rightarrow s_{12}^2$ in \eqn{H-identity}.

\section{Six loop integrand construction}
With the X and H identities established, we set out to demonstrate their usefulness  by constructing the complete six-loop four-point integrand in $\mathcal{N}=4$ SYM. To assemble the integrand from cuts, we follow the generalized unitarity method
\cite{Bern:1994zx,Bern:1994cg,Bern:1997sc,Bern:2004ky,Britto:2004nc} in the refined form known as the method of maximal cuts
\cite{Bern:2007ct,Bern:2008pv,Bern:2010tq,Bern:2012uf,Bern:2018jmv}. 

The construction can be parsed into three steps.
The first step is to enumerate all candidate diagrams entering the six-loop integrand, and grade them by the \emph{cut level} $k$. Cut level zero corresponds to the maximal-cut diagrams, obtained by attaching four external legs to the edges of all cubic six-loop vacuum diagrams, which are known~\cite{BUSSEMAKE1977234}. By the ${\cal N}=4$ SYM ``no-triangle rule''~\cite{Bjerrum-Bohr:2006xbk}, we exclude one-loop triangles, bubbles, or tadpoles subdiagrams, as their maximal cuts are zero.
Cut level $k$ diagrams are then obtained by contracting an internal edge (by merging the edge's vertices) in the cut level $(k-1)$ diagrams, in all possible ways, modding out by graph isomorphisms. (If an edge only connects to one vertex, i.e. tadpole, contraction is avoided.)
This recursively constructs the (next-to){}$^k$-maximal-cut (\nkmc) diagrams, and a typical such graph is below called $\gamma^{(k)}$. The final counts are given in Table~\ref{tab:six-data}, first row.

\begin{table}[ht]
  \begin{center}
    \begin{tabular}{c|ccccc|c}
      \nkmc & 0 & 1 & 2 & 3 & 4&$\sum$ \\ \hline
      all cuts & 5548 & 41649 &156853 &363963 &576582 &1144595\\
      box cuts & 5218 & 38721 & 144428 & 333501 & 526082 & 1047950 \\
      $n_{\gamma^{(k)}}\neq 0$ &4420 &16776 &37653 &56717 &36087 &151653
    \end{tabular}
    \caption{Counts for candidate cuts: (a) all generated from cubic no-triangle graphs; (b) cross-checked with box cuts~\cite{Carrasco:2011hw,Bern:2010tq,Bern:2012uf}; (c) require non-vanishing numerator contribution.}
    \label{tab:six-data}
  \end{center}
\end{table}

The second step is to assign physical expressions to each diagram, specifically two non-trivial expressions for each graph: a diagram's \emph{cut} $\cgamma$ is a rational and gauge invariant (unique) function that corresponds to a physical factorization process in the theory; a diagram's \emph{numerator} $n_{\gamma^{(k)}}$ is a non-unique (gauge dependent) polynomial that describes local interactions similar to numerators of Feynman diagrams. The unique $\cgamma$ are determined by the X and H identities. 
From the numerators, the amplitude is
\begin{equation}
    \mathcal{A}_4^{L=6} = i^6g_{\rm YM}^{14} {\cal K} \sum_{k=0}^4 \sum_{\,\gamma^{(k)}\,} \int \frac{\text{d}^{6D} \ell}{(2\pi)^{6D}} \frac{n_{\gamma^{(k)}}}{S_{\gamma^{(k)}} d_{\gamma^{(k)}}}\,
\end{equation}
where $S_{\gamma^{(k)}}$ and $d_{\gamma^{(k)}}$ are a graph's symmetry factor and product of propagator denominators, respectively.  ${\cal K}$ is a crossing-symmetric kinematic factor that captures the external state dependence, ${\cal K}= s_{12} s_{23} A_{\rm SYM}^{\rm tree}(1,2,3,4)$. Because the first sum runs over both cubic $(k=0)$ and contact $(k>0)$ diagrams, it is convenient to make the numerators $n_{\gamma^{(k)}}$ functions of both kinematic and color data, hence we use no separate color factors. Similar to previous constructions of the $L$-loop four-point amplitude in ${\cal N}=4$ SYM, the cut level $k=L-2$ is the last one that contains new data, hence we anticipate no contributing diagrams beyond $k>4$ (also confirmed by our cuts). 

In the third step, we construct the $n_{\gamma^{(k)}}$ by appropriately matching to the unique $\cgamma$, on the kinematic support where every factor in $d_{\gamma^{(k)}}$ vanish, denoted by $d_{\gamma^{(0)}}\rightarrow 0$.
For maximal cuts, the matching is simply
\begin{equation}
    n_{\gamma^{(0)}} \big|_{d_{\gamma^{(0)}}\rightarrow 0} = \mathcal{C}_{\gamma^{(0)}}
\end{equation}
The matching procedure on a \nkmc[k]
diagram $\gamma^{(k)}$ can be expressed as (with $d_{\gamma^{(k)}} \rightarrow  0$ implicit from now on)
\begin{equation}
  \locp = \cgamma-\ratr
  \label{eq:ans-def}\,,
\end{equation}
where $\cgamma$ is the unique cut we need to match, and
$\ratr$ is a rational function obtained by summing over previously-determined lower-$k$ diagrams (numerator over non-vanishing denominator factors) that shares the same poles as $\cgamma$. 
By construction
$\ratr$ matches all lower-$k$ cuts,
hence $\locp$ must be a local polynomial. We would like to identify $\locp$ with the numerator $n_{\gamma^{(k)}}$; however, first we must promote $\locp$ to an off-shell quantity respecting the automorphism symmetry of the graph $\gamma^{(k)}$ (recall that $S_{\gamma^{(k)}}$ equals the order of the automorphism group). This can be done in several ways, such as explicit symmetrization, or using an Ansatz. Here we use an approach that directly identifies special combinations of generalized Mandelstam variables that are automorphism invariants~\cite{AEJPM}. Hence we can promote $\locp \rightarrow n_{\gamma^{(k)}}$.

Let us return to the new identities. The \xid{} directly supplies $\cgamma$ expressions from higher-loop planar integrands, currently known explicitly up to $L=10$ \cite{Bourjaily:2016evz}. However, in the following we use the \hid, and show that it alone is sufficiently powerful to determine the complete six-loop integrand. Recall that the \hid{} does not directly calculate $\cgamma$, but rather provide a
set of constraints,
\begin{equation}
  \lim_{\ell_m \to 0} \locp =
2 \ell_i {\cdot} \ell_j\,\mathcal{C}_{\gamma^{(k)}\setminus \ell_m} \,-\, \lim_{\ell_m \to 0} \ratr \,,
  \label{eq:h-impl}
\end{equation}
where $\mathcal{C}_{\gamma^{(k)}\setminus \ell_m}$ is a known \emph{lower}-loop cut with edge $\ell_m$ removed, and its prefactor $2 \ell_i \cdot \ell_j$ corresponds to the momenta of the two edges that $\ell_m$ was attached to. Consider a pictorial example: with a slight 
shuffe, \eqn{eq:h-impl} becomes
\begin{align}
   &\lim_{\ell_m \to 0}\left( \locp + \ratr \right) =2 \ell_i {\cdot} \ell_j\,\mathcal{C}_{\gamma^{(k)}\setminus \ell_m} \notag\\
  \Rightarrow& \arrdiag{0.12\textwidth}{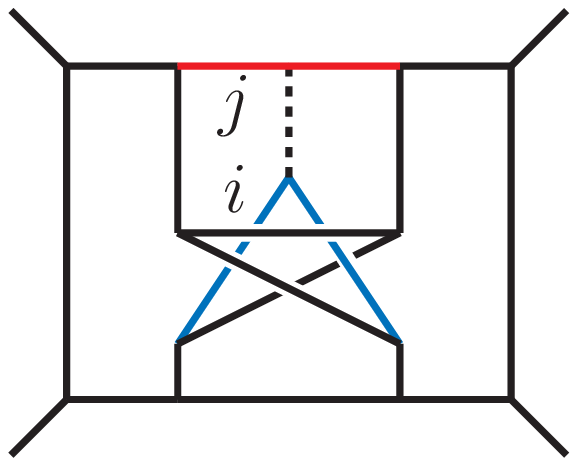} = 2 \ell_i \cdot \ell_j
  \arrdiag{0.12\textwidth}{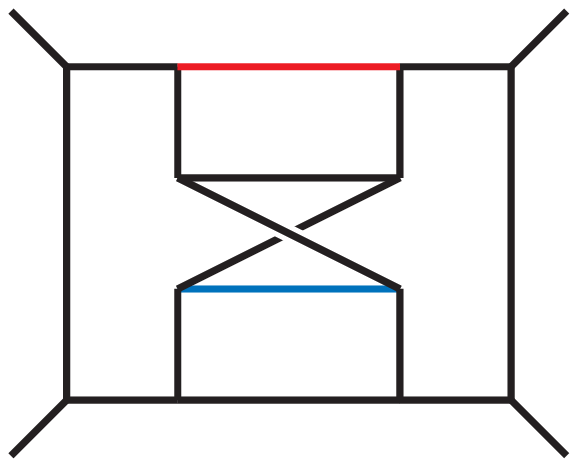}
\end{align}
where the pictures are representative 6- and 5-loop diagrams producing the functions $\locp+\ratr$ and $\mathcal{C}_{\gamma^{(k)}\setminus \ell_m}$, respectively.
The edges labeled by $i,j$ are on-shell, and the momentum of the dashed line vanishes, $\ell_m=0$.  Repeatedly applying this
procedure to every edge of $\gamma^{(k)}$ provides all of the information required to uniquely determine $\locp$. Note that, if the lower-loop cut happens to vanish, \eqn{eq:h-impl} still provides valuable information,  in a similar vein to how vanishing cuts were used to triangulate the Amplituhedron in Ref.~\cite{Arkani-Hamed:2014dca}.  

In the six-loop construction, only mild additional power-counting assumptions on the integrand are required to fully constrain the numerators.
For each candidate diagram we use \eqn{eq:h-impl} to construct a $\locp$ ($n_{\gamma^{(k)}}$ off shell) that satisfies: 1) every monomial term is proportional to either $s_{12}$ or $s_{23}$; 2) if there are any terms left unconstrained that violate the no-triangle power counting, we 
safely drop them.  Condition 1 holds for all lower-loop integrands
\cite{Bern:2010ue,Bern:2012uf,Bern:2017ucb} as well as in the planar
six-loop amplitude~\cite{Bern:2012di}.  Condition 2 is needed because there are instances of particularly simple ladder-type diagrams that are not uniquely constrained by the \hid, but subdiagram power counting rules out such freedom. We defer an in-depth discussion of the implementation of the method and details to upcoming work~\cite{Carrasco:2022zzz}.  Both assumptions are validated by the
consistency checks of the final integrand.  The third row of
Table~\ref{tab:six-data} provides the final count of non-zero diagrams that contribute to the integrand.
The complete ${\cal N}=4$ SYM six-loop integrand, with explicit numerators for all diagram topologies, is provided in
Ref.~\cite{SixIntZenodo}, including additional user-manual details.

By construction, the maximal cut method gives an integrand that satisfies all independent unitarity cuts. However, cross-checks are always desirable. We have subjected our six-loop integrand to two independent consistency checks. 
The first is verifying that it correctly produces all so-called generalized box cuts~\cite{Carrasco:2011hw,Bern:2010tq,Bern:2012uf}. This
class of cuts targets graphs with lower-loop ($6>L\ge1$) four-point sub-diagrams, which are factorized into the product of two known four-point cuts, one at $L$ loops and one at $6{-}L$ loops.  Cuts verified with this method constitute a large proportion of each cut level $k$, as shown in the second row of
Table~\ref{tab:six-data}.  
Second, we have verified that all contact-contributing non-multigraph 
\nkmc[4] are consistent with the $n$-point BCJ amplitude relations~\cite{Bern:2008qj}, as applied to sub-amplitudes within the cuts. This second check is a highly nontrivial verification that no diagrams or terms were lost in each cut, as all contributions must delicately conspire between different $k\le4$ diagrams to satisfy the relations.

\section{Conclusions}
In this Letter, we have presented two new identities which have extensive applications in the recursive
calculation of unitarity cuts in many theories (including QCD), avoiding laborious state sums and manifesting desirable symmetries.  
We demonstrated the power of the \hid{} by using it alone to compute the four-point six-loop $D$-dimensional integrand in $\mathcal{N}=4$ SYM, providing the first non-planar result at this loop order (see refs.~\cite{Bourjaily:2011hi,Eden:2012tu,Bern:2012di} for planar results). Maximal SYM has a rich integrand structure both in the planar~\cite{Bourjaily:2011hi,
Golden:2012hi,
Bourjaily:2015jna,
Bourjaily:2016evz,
Bourjaily:2017wjl,
Arkani-Hamed:2018rsk,
Langer:2019iuo}
and non-planar~\cite{Mafra:2015mja,
Bourjaily:2019iqr,
Bourjaily:2019gqu} cases, with analogous multiloop string perspectives~\cite{Gomez:2013sla,Geyer:2016wjx,Geyer:2018xwu,Geyer:2019hnn,DHoker:2020prr, Geyer:2021oox}, and we provide here an ample source for further studies~\cite{SixIntZenodo}. 

The current result constitutes major progress towards determining the UV behavior of $\mathcal{N}=8$ supergravity at seven loops.
However, before getting there further six-loop studies are motivated.
First, the current understanding of (generalized) double-copy construction starts from a cubic integrand  representation~\cite{Bern:2017yxu}.
Second, previous studies~\cite{Bern:2018jmv,Bern:2017ucb,Bern:2017yxu,Bern:2012uf,Bern:2009kd,Bern:2008pv} have demonstrated the significant simplification of gravity calculations that comes from using a gauge theory integrand that manifests the expected sub-diagram power counting. Thus, the current integrand can be improved upon by efficiently constructing a cubic representation, perhaps with improved manifest sub-diagram power counting.

The six-loop integrand provides an essential stepping stone towards the corresponding seven-loop SYM calculation, using the new recursive methods presented in this Letter. 
The combinatorial growth of complexity when reaching seven loops will demand significantly streamlined tools, and we anticipate that the new identities will play a key role.

\paragraph{Acknowledgements:}
We thank Zvi Bern, Radu Roiban, and Oliver Schlotterer for useful  discussions, and  Bram Verbeek and Julio Parra-Martinez for collaborations on related topics.
This work was supported by the DOE under contract DE-SC0021485 and by the Alfred P. Sloan Foundation, 
the Knut and Alice Wallenberg Foundation under grants KAW 2018.0116 ({\it From Scattering Amplitudes to Gravitational Waves}) and KAW 2018.0162, the Swedish Research Council under grant 621-2014-5722, and the Ragnar S\"{o}derberg Foundation (Swedish Foundations' Starting Grant).

\bibliography{six_symBib}

\begin{thebibliography}{77}%
\makeatletter
\providecommand \@ifxundefined [1]{%
 \@ifx{#1\undefined}
}%
\providecommand \@ifnum [1]{%
 \ifnum #1\expandafter \@firstoftwo
 \else \expandafter \@secondoftwo
 \fi
}%
\providecommand \@ifx [1]{%
 \ifx #1\expandafter \@firstoftwo
 \else \expandafter \@secondoftwo
 \fi
}%
\providecommand \natexlab [1]{#1}%
\providecommand \enquote  [1]{``#1''}%
\providecommand \bibnamefont  [1]{#1}%
\providecommand \bibfnamefont [1]{#1}%
\providecommand \citenamefont [1]{#1}%
\providecommand \href@noop [0]{\@secondoftwo}%
\providecommand \href [0]{\begingroup \@sanitize@url \@href}%
\providecommand \@href[1]{\@@startlink{#1}\@@href}%
\providecommand \@@href[1]{\endgroup#1\@@endlink}%
\providecommand \@sanitize@url [0]{\catcode `\\12\catcode `\$12\catcode
  `\&12\catcode `\#12\catcode `\^12\catcode `\_12\catcode `\%12\relax}%
\providecommand \@@startlink[1]{}%
\providecommand \@@endlink[0]{}%
\providecommand \url  [0]{\begingroup\@sanitize@url \@url }%
\providecommand \@url [1]{\endgroup\@href {#1}{\urlprefix }}%
\providecommand \urlprefix  [0]{URL }%
\providecommand \Eprint [0]{\href }%
\providecommand \doibase [0]{http://dx.doi.org/}%
\providecommand \selectlanguage [0]{\@gobble}%
\providecommand \bibinfo  [0]{\@secondoftwo}%
\providecommand \bibfield  [0]{\@secondoftwo}%
\providecommand \translation [1]{[#1]}%
\providecommand \BibitemOpen [0]{}%
\providecommand \bibitemStop [0]{}%
\providecommand \bibitemNoStop [0]{.\EOS\space}%
\providecommand \EOS [0]{\spacefactor3000\relax}%
\providecommand \BibitemShut  [1]{\csname bibitem#1\endcsname}%
\let\auto@bib@innerbib\@empty
\bibitem [{\citenamefont {Goroff}\ and\ \citenamefont
  {Sagnotti}(1986)}]{Goroff:1985th}%
  \BibitemOpen
  \bibfield  {author} {\bibinfo {author} {\bibfnamefont {M.~H.}\ \bibnamefont
  {Goroff}}\ and\ \bibinfo {author} {\bibfnamefont {A.}~\bibnamefont
  {Sagnotti}},\ }\href {\doibase 10.1016/0550-3213(86)90193-8} {\bibfield
  {journal} {\bibinfo  {journal} {Nucl. Phys. B}\ }\textbf {\bibinfo {volume}
  {266}},\ \bibinfo {pages} {709} (\bibinfo {year} {1986})}\BibitemShut
  {NoStop}%
\bibitem [{\citenamefont {Cremmer}\ \emph {et~al.}(1978)\citenamefont
  {Cremmer}, \citenamefont {Julia},\ and\ \citenamefont
  {Scherk}}]{Cremmer:1978km}%
  \BibitemOpen
  \bibfield  {author} {\bibinfo {author} {\bibfnamefont {E.}~\bibnamefont
  {Cremmer}}, \bibinfo {author} {\bibfnamefont {B.}~\bibnamefont {Julia}}, \
  and\ \bibinfo {author} {\bibfnamefont {J.}~\bibnamefont {Scherk}},\ }\href
  {\doibase 10.1016/0370-2693(78)90894-8} {\bibfield  {journal} {\bibinfo
  {journal} {Phys. Lett. B}\ }\textbf {\bibinfo {volume} {76}},\ \bibinfo
  {pages} {409} (\bibinfo {year} {1978})}\BibitemShut {NoStop}%
\bibitem [{\citenamefont {Kallosh}(2009)}]{Kallosh:2009db}%
  \BibitemOpen
  \bibfield  {author} {\bibinfo {author} {\bibfnamefont {R.}~\bibnamefont
  {Kallosh}},\ }\href {\doibase 10.1103/PhysRevD.80.105022} {\bibfield
  {journal} {\bibinfo  {journal} {Phys. Rev. D}\ }\textbf {\bibinfo {volume}
  {80}},\ \bibinfo {pages} {105022} (\bibinfo {year} {2009})},\ \Eprint
  {http://arxiv.org/abs/0903.4630} {arXiv:0903.4630 [hep-th]} \BibitemShut
  {NoStop}%
\bibitem [{\citenamefont {Bossard}\ \emph {et~al.}(2009)\citenamefont
  {Bossard}, \citenamefont {Howe},\ and\ \citenamefont
  {Stelle}}]{Bossard:2009sy}%
  \BibitemOpen
  \bibfield  {author} {\bibinfo {author} {\bibfnamefont {G.}~\bibnamefont
  {Bossard}}, \bibinfo {author} {\bibfnamefont {P.~S.}\ \bibnamefont {Howe}}, \
  and\ \bibinfo {author} {\bibfnamefont {K.~S.}\ \bibnamefont {Stelle}},\
  }\href {\doibase 10.1007/s10714-009-0775-0} {\bibfield  {journal} {\bibinfo
  {journal} {Gen. Rel. Grav.}\ }\textbf {\bibinfo {volume} {41}},\ \bibinfo
  {pages} {919} (\bibinfo {year} {2009})},\ \Eprint
  {http://arxiv.org/abs/0901.4661} {arXiv:0901.4661 [hep-th]} \BibitemShut
  {NoStop}%
\bibitem [{\citenamefont {Howe}\ and\ \citenamefont
  {Lindstrom}(1981)}]{Howe:1980th}%
  \BibitemOpen
  \bibfield  {author} {\bibinfo {author} {\bibfnamefont {P.~S.}\ \bibnamefont
  {Howe}}\ and\ \bibinfo {author} {\bibfnamefont {U.}~\bibnamefont
  {Lindstrom}},\ }\href {\doibase 10.1016/0550-3213(81)90537-X} {\bibfield
  {journal} {\bibinfo  {journal} {Nucl. Phys. B}\ }\textbf {\bibinfo {volume}
  {181}},\ \bibinfo {pages} {487} (\bibinfo {year} {1981})}\BibitemShut
  {NoStop}%
\bibitem [{\citenamefont {Kallosh}(1981)}]{Kallosh:1980fi}%
  \BibitemOpen
  \bibfield  {author} {\bibinfo {author} {\bibfnamefont {R.~E.}\ \bibnamefont
  {Kallosh}},\ }\href {\doibase 10.1016/0370-2693(81)90964-3} {\bibfield
  {journal} {\bibinfo  {journal} {Phys. Lett. B}\ }\textbf {\bibinfo {volume}
  {99}},\ \bibinfo {pages} {122} (\bibinfo {year} {1981})}\BibitemShut
  {NoStop}%
\bibitem [{\citenamefont {Berkovits}(2007)}]{Berkovits:2006vc}%
  \BibitemOpen
  \bibfield  {author} {\bibinfo {author} {\bibfnamefont {N.}~\bibnamefont
  {Berkovits}},\ }\href {\doibase 10.1103/PhysRevLett.98.211601} {\bibfield
  {journal} {\bibinfo  {journal} {Phys. Rev. Lett.}\ }\textbf {\bibinfo
  {volume} {98}},\ \bibinfo {pages} {211601} (\bibinfo {year} {2007})},\
  \Eprint {http://arxiv.org/abs/hep-th/0609006} {arXiv:hep-th/0609006}
  \BibitemShut {NoStop}%
\bibitem [{\citenamefont {Green}\ \emph
  {et~al.}(2007{\natexlab{a}})\citenamefont {Green}, \citenamefont {Russo},\
  and\ \citenamefont {Vanhove}}]{Green:2006yu}%
  \BibitemOpen
  \bibfield  {author} {\bibinfo {author} {\bibfnamefont {M.~B.}\ \bibnamefont
  {Green}}, \bibinfo {author} {\bibfnamefont {J.~G.}\ \bibnamefont {Russo}}, \
  and\ \bibinfo {author} {\bibfnamefont {P.}~\bibnamefont {Vanhove}},\ }\href
  {\doibase 10.1103/PhysRevLett.98.131602} {\bibfield  {journal} {\bibinfo
  {journal} {Phys. Rev. Lett.}\ }\textbf {\bibinfo {volume} {98}},\ \bibinfo
  {pages} {131602} (\bibinfo {year} {2007}{\natexlab{a}})},\ \Eprint
  {http://arxiv.org/abs/hep-th/0611273} {arXiv:hep-th/0611273} \BibitemShut
  {NoStop}%
\bibitem [{\citenamefont {Green}\ \emph
  {et~al.}(2007{\natexlab{b}})\citenamefont {Green}, \citenamefont {Russo},\
  and\ \citenamefont {Vanhove}}]{Green:2006gt}%
  \BibitemOpen
  \bibfield  {author} {\bibinfo {author} {\bibfnamefont {M.~B.}\ \bibnamefont
  {Green}}, \bibinfo {author} {\bibfnamefont {J.~G.}\ \bibnamefont {Russo}}, \
  and\ \bibinfo {author} {\bibfnamefont {P.}~\bibnamefont {Vanhove}},\ }\href
  {\doibase 10.1088/1126-6708/2007/02/099} {\bibfield  {journal} {\bibinfo
  {journal} {JHEP}\ }\textbf {\bibinfo {volume} {02}},\ \bibinfo {pages} {099}
  (\bibinfo {year} {2007}{\natexlab{b}})},\ \Eprint
  {http://arxiv.org/abs/hep-th/0610299} {arXiv:hep-th/0610299} \BibitemShut
  {NoStop}%
\bibitem [{\citenamefont {Marcus}\ and\ \citenamefont
  {Sagnotti}(1985)}]{Marcus:1984ei}%
  \BibitemOpen
  \bibfield  {author} {\bibinfo {author} {\bibfnamefont {N.}~\bibnamefont
  {Marcus}}\ and\ \bibinfo {author} {\bibfnamefont {A.}~\bibnamefont
  {Sagnotti}},\ }\href {\doibase 10.1016/0550-3213(85)90386-4} {\bibfield
  {journal} {\bibinfo  {journal} {Nucl. Phys. B}\ }\textbf {\bibinfo {volume}
  {256}},\ \bibinfo {pages} {77} (\bibinfo {year} {1985})}\BibitemShut
  {NoStop}%
\bibitem [{\citenamefont {Berkovits}\ \emph {et~al.}(2009)\citenamefont
  {Berkovits}, \citenamefont {Green}, \citenamefont {Russo},\ and\
  \citenamefont {Vanhove}}]{Berkovits:2009aw}%
  \BibitemOpen
  \bibfield  {author} {\bibinfo {author} {\bibfnamefont {N.}~\bibnamefont
  {Berkovits}}, \bibinfo {author} {\bibfnamefont {M.~B.}\ \bibnamefont
  {Green}}, \bibinfo {author} {\bibfnamefont {J.~G.}\ \bibnamefont {Russo}}, \
  and\ \bibinfo {author} {\bibfnamefont {P.}~\bibnamefont {Vanhove}},\ }\href
  {\doibase 10.1088/1126-6708/2009/11/063} {\bibfield  {journal} {\bibinfo
  {journal} {JHEP}\ }\textbf {\bibinfo {volume} {11}},\ \bibinfo {pages} {063}
  (\bibinfo {year} {2009})},\ \Eprint {http://arxiv.org/abs/0908.1923}
  {arXiv:0908.1923 [hep-th]} \BibitemShut {NoStop}%
\bibitem [{\citenamefont {Vanhove}(2010)}]{Vanhove:2010nf}%
  \BibitemOpen
  \bibfield  {author} {\bibinfo {author} {\bibfnamefont {P.}~\bibnamefont
  {Vanhove}},\ }\href@noop {} {\  (\bibinfo {year} {2010})},\ \Eprint
  {http://arxiv.org/abs/1004.1392} {arXiv:1004.1392 [hep-th]} \BibitemShut
  {NoStop}%
\bibitem [{\citenamefont {Beisert}\ \emph {et~al.}(2011)\citenamefont
  {Beisert}, \citenamefont {Elvang}, \citenamefont {Freedman}, \citenamefont
  {Kiermaier}, \citenamefont {Morales},\ and\ \citenamefont
  {Stieberger}}]{Beisert:2010jx}%
  \BibitemOpen
  \bibfield  {author} {\bibinfo {author} {\bibfnamefont {N.}~\bibnamefont
  {Beisert}}, \bibinfo {author} {\bibfnamefont {H.}~\bibnamefont {Elvang}},
  \bibinfo {author} {\bibfnamefont {D.~Z.}\ \bibnamefont {Freedman}}, \bibinfo
  {author} {\bibfnamefont {M.}~\bibnamefont {Kiermaier}}, \bibinfo {author}
  {\bibfnamefont {A.}~\bibnamefont {Morales}}, \ and\ \bibinfo {author}
  {\bibfnamefont {S.}~\bibnamefont {Stieberger}},\ }\href {\doibase
  10.1016/j.physletb.2010.09.069} {\bibfield  {journal} {\bibinfo  {journal}
  {Phys. Lett. B}\ }\textbf {\bibinfo {volume} {694}},\ \bibinfo {pages} {265}
  (\bibinfo {year} {2011})},\ \Eprint {http://arxiv.org/abs/1009.1643}
  {arXiv:1009.1643 [hep-th]} \BibitemShut {NoStop}%
\bibitem [{\citenamefont {Bossard}\ \emph
  {et~al.}(2011{\natexlab{a}})\citenamefont {Bossard}, \citenamefont {Howe},\
  and\ \citenamefont {Stelle}}]{Bossard:2010bd}%
  \BibitemOpen
  \bibfield  {author} {\bibinfo {author} {\bibfnamefont {G.}~\bibnamefont
  {Bossard}}, \bibinfo {author} {\bibfnamefont {P.~S.}\ \bibnamefont {Howe}}, \
  and\ \bibinfo {author} {\bibfnamefont {K.~S.}\ \bibnamefont {Stelle}},\
  }\href {\doibase 10.1007/JHEP01(2011)020} {\bibfield  {journal} {\bibinfo
  {journal} {JHEP}\ }\textbf {\bibinfo {volume} {01}},\ \bibinfo {pages} {020}
  (\bibinfo {year} {2011}{\natexlab{a}})},\ \Eprint
  {http://arxiv.org/abs/1009.0743} {arXiv:1009.0743 [hep-th]} \BibitemShut
  {NoStop}%
\bibitem [{\citenamefont {Bjornsson}(2011)}]{Bjornsson:2010wu}%
  \BibitemOpen
  \bibfield  {author} {\bibinfo {author} {\bibfnamefont {J.}~\bibnamefont
  {Bjornsson}},\ }\href {\doibase 10.1007/JHEP01(2011)002} {\bibfield
  {journal} {\bibinfo  {journal} {JHEP}\ }\textbf {\bibinfo {volume} {01}},\
  \bibinfo {pages} {002} (\bibinfo {year} {2011})},\ \Eprint
  {http://arxiv.org/abs/1009.5906} {arXiv:1009.5906 [hep-th]} \BibitemShut
  {NoStop}%
\bibitem [{\citenamefont {Bossard}\ \emph
  {et~al.}(2011{\natexlab{b}})\citenamefont {Bossard}, \citenamefont {Howe},
  \citenamefont {Stelle},\ and\ \citenamefont {Vanhove}}]{Bossard:2011tq}%
  \BibitemOpen
  \bibfield  {author} {\bibinfo {author} {\bibfnamefont {G.}~\bibnamefont
  {Bossard}}, \bibinfo {author} {\bibfnamefont {P.~S.}\ \bibnamefont {Howe}},
  \bibinfo {author} {\bibfnamefont {K.~S.}\ \bibnamefont {Stelle}}, \ and\
  \bibinfo {author} {\bibfnamefont {P.}~\bibnamefont {Vanhove}},\ }\href
  {\doibase 10.1088/0264-9381/28/21/215005} {\bibfield  {journal} {\bibinfo
  {journal} {Class. Quant. Grav.}\ }\textbf {\bibinfo {volume} {28}},\ \bibinfo
  {pages} {215005} (\bibinfo {year} {2011}{\natexlab{b}})},\ \Eprint
  {http://arxiv.org/abs/1105.6087} {arXiv:1105.6087 [hep-th]} \BibitemShut
  {NoStop}%
\bibitem [{\citenamefont {Brink}\ \emph {et~al.}(1977)\citenamefont {Brink},
  \citenamefont {Schwarz},\ and\ \citenamefont {Scherk}}]{Brink:1976bc}%
  \BibitemOpen
  \bibfield  {author} {\bibinfo {author} {\bibfnamefont {L.}~\bibnamefont
  {Brink}}, \bibinfo {author} {\bibfnamefont {J.~H.}\ \bibnamefont {Schwarz}},
  \ and\ \bibinfo {author} {\bibfnamefont {J.}~\bibnamefont {Scherk}},\ }\href
  {\doibase 10.1016/0550-3213(77)90328-5} {\bibfield  {journal} {\bibinfo
  {journal} {Nucl. Phys. B}\ }\textbf {\bibinfo {volume} {121}},\ \bibinfo
  {pages} {77} (\bibinfo {year} {1977})}\BibitemShut {NoStop}%
\bibitem [{\citenamefont {Kawai}\ \emph {et~al.}(1986)\citenamefont {Kawai},
  \citenamefont {Lewellen},\ and\ \citenamefont {Tye}}]{Kawai:1985xq}%
  \BibitemOpen
  \bibfield  {author} {\bibinfo {author} {\bibfnamefont {H.}~\bibnamefont
  {Kawai}}, \bibinfo {author} {\bibfnamefont {D.~C.}\ \bibnamefont {Lewellen}},
  \ and\ \bibinfo {author} {\bibfnamefont {S.~H.~H.}\ \bibnamefont {Tye}},\
  }\href {\doibase 10.1016/0550-3213(86)90362-7} {\bibfield  {journal}
  {\bibinfo  {journal} {Nucl. Phys. B}\ }\textbf {\bibinfo {volume} {269}},\
  \bibinfo {pages} {1} (\bibinfo {year} {1986})}\BibitemShut {NoStop}%
\bibitem [{\citenamefont {Bern}\ \emph
  {et~al.}(2008{\natexlab{a}})\citenamefont {Bern}, \citenamefont {Carrasco},\
  and\ \citenamefont {Johansson}}]{Bern:2008qj}%
  \BibitemOpen
  \bibfield  {author} {\bibinfo {author} {\bibfnamefont {Z.}~\bibnamefont
  {Bern}}, \bibinfo {author} {\bibfnamefont {J.~J.~M.}\ \bibnamefont
  {Carrasco}}, \ and\ \bibinfo {author} {\bibfnamefont {H.}~\bibnamefont
  {Johansson}},\ }\href {\doibase 10.1103/PhysRevD.78.085011} {\bibfield
  {journal} {\bibinfo  {journal} {Phys. Rev. D}\ }\textbf {\bibinfo {volume}
  {78}},\ \bibinfo {pages} {085011} (\bibinfo {year} {2008}{\natexlab{a}})},\
  \Eprint {http://arxiv.org/abs/0805.3993} {arXiv:0805.3993 [hep-ph]}
  \BibitemShut {NoStop}%
\bibitem [{\citenamefont {Bern}\ \emph
  {et~al.}(2010{\natexlab{a}})\citenamefont {Bern}, \citenamefont {Carrasco},\
  and\ \citenamefont {Johansson}}]{Bern:2010ue}%
  \BibitemOpen
  \bibfield  {author} {\bibinfo {author} {\bibfnamefont {Z.}~\bibnamefont
  {Bern}}, \bibinfo {author} {\bibfnamefont {J.~J.~M.}\ \bibnamefont
  {Carrasco}}, \ and\ \bibinfo {author} {\bibfnamefont {H.}~\bibnamefont
  {Johansson}},\ }\href {\doibase 10.1103/PhysRevLett.105.061602} {\bibfield
  {journal} {\bibinfo  {journal} {Phys. Rev. Lett.}\ }\textbf {\bibinfo
  {volume} {105}},\ \bibinfo {pages} {061602} (\bibinfo {year}
  {2010}{\natexlab{a}})},\ \Eprint {http://arxiv.org/abs/1004.0476}
  {arXiv:1004.0476 [hep-th]} \BibitemShut {NoStop}%
\bibitem [{\citenamefont {Bern}\ \emph {et~al.}(1994)\citenamefont {Bern},
  \citenamefont {Dixon}, \citenamefont {Dunbar},\ and\ \citenamefont
  {Kosower}}]{Bern:1994zx}%
  \BibitemOpen
  \bibfield  {author} {\bibinfo {author} {\bibfnamefont {Z.}~\bibnamefont
  {Bern}}, \bibinfo {author} {\bibfnamefont {L.~J.}\ \bibnamefont {Dixon}},
  \bibinfo {author} {\bibfnamefont {D.~C.}\ \bibnamefont {Dunbar}}, \ and\
  \bibinfo {author} {\bibfnamefont {D.~A.}\ \bibnamefont {Kosower}},\ }\href
  {\doibase 10.1016/0550-3213(94)90179-1} {\bibfield  {journal} {\bibinfo
  {journal} {Nucl. Phys. B}\ }\textbf {\bibinfo {volume} {425}},\ \bibinfo
  {pages} {217} (\bibinfo {year} {1994})},\ \Eprint
  {http://arxiv.org/abs/hep-ph/9403226} {arXiv:hep-ph/9403226} \BibitemShut
  {NoStop}%
\bibitem [{\citenamefont {Bern}\ \emph {et~al.}(1995)\citenamefont {Bern},
  \citenamefont {Dixon}, \citenamefont {Dunbar},\ and\ \citenamefont
  {Kosower}}]{Bern:1994cg}%
  \BibitemOpen
  \bibfield  {author} {\bibinfo {author} {\bibfnamefont {Z.}~\bibnamefont
  {Bern}}, \bibinfo {author} {\bibfnamefont {L.~J.}\ \bibnamefont {Dixon}},
  \bibinfo {author} {\bibfnamefont {D.~C.}\ \bibnamefont {Dunbar}}, \ and\
  \bibinfo {author} {\bibfnamefont {D.~A.}\ \bibnamefont {Kosower}},\ }\href
  {\doibase 10.1016/0550-3213(94)00488-Z} {\bibfield  {journal} {\bibinfo
  {journal} {Nucl. Phys. B}\ }\textbf {\bibinfo {volume} {435}},\ \bibinfo
  {pages} {59} (\bibinfo {year} {1995})},\ \Eprint
  {http://arxiv.org/abs/hep-ph/9409265} {arXiv:hep-ph/9409265} \BibitemShut
  {NoStop}%
\bibitem [{\citenamefont {Bern}\ \emph
  {et~al.}(2007{\natexlab{a}})\citenamefont {Bern}, \citenamefont {Carrasco},
  \citenamefont {Dixon}, \citenamefont {Johansson}, \citenamefont {Kosower},\
  and\ \citenamefont {Roiban}}]{Bern:2007hh}%
  \BibitemOpen
  \bibfield  {author} {\bibinfo {author} {\bibfnamefont {Z.}~\bibnamefont
  {Bern}}, \bibinfo {author} {\bibfnamefont {J.~J.}\ \bibnamefont {Carrasco}},
  \bibinfo {author} {\bibfnamefont {L.~J.}\ \bibnamefont {Dixon}}, \bibinfo
  {author} {\bibfnamefont {H.}~\bibnamefont {Johansson}}, \bibinfo {author}
  {\bibfnamefont {D.~A.}\ \bibnamefont {Kosower}}, \ and\ \bibinfo {author}
  {\bibfnamefont {R.}~\bibnamefont {Roiban}},\ }\href {\doibase
  10.1103/PhysRevLett.98.161303} {\bibfield  {journal} {\bibinfo  {journal}
  {Phys. Rev. Lett.}\ }\textbf {\bibinfo {volume} {98}},\ \bibinfo {pages}
  {161303} (\bibinfo {year} {2007}{\natexlab{a}})},\ \Eprint
  {http://arxiv.org/abs/hep-th/0702112} {arXiv:hep-th/0702112} \BibitemShut
  {NoStop}%
\bibitem [{\citenamefont {Bern}\ \emph
  {et~al.}(2008{\natexlab{b}})\citenamefont {Bern}, \citenamefont {Carrasco},
  \citenamefont {Dixon}, \citenamefont {Johansson},\ and\ \citenamefont
  {Roiban}}]{Bern:2008pv}%
  \BibitemOpen
  \bibfield  {author} {\bibinfo {author} {\bibfnamefont {Z.}~\bibnamefont
  {Bern}}, \bibinfo {author} {\bibfnamefont {J.~J.~M.}\ \bibnamefont
  {Carrasco}}, \bibinfo {author} {\bibfnamefont {L.~J.}\ \bibnamefont {Dixon}},
  \bibinfo {author} {\bibfnamefont {H.}~\bibnamefont {Johansson}}, \ and\
  \bibinfo {author} {\bibfnamefont {R.}~\bibnamefont {Roiban}},\ }\href
  {\doibase 10.1103/PhysRevD.78.105019} {\bibfield  {journal} {\bibinfo
  {journal} {Phys. Rev. D}\ }\textbf {\bibinfo {volume} {78}},\ \bibinfo
  {pages} {105019} (\bibinfo {year} {2008}{\natexlab{b}})},\ \Eprint
  {http://arxiv.org/abs/0808.4112} {arXiv:0808.4112 [hep-th]} \BibitemShut
  {NoStop}%
\bibitem [{\citenamefont {Bern}\ \emph {et~al.}(2009)\citenamefont {Bern},
  \citenamefont {Carrasco}, \citenamefont {Dixon}, \citenamefont {Johansson},\
  and\ \citenamefont {Roiban}}]{Bern:2009kd}%
  \BibitemOpen
  \bibfield  {author} {\bibinfo {author} {\bibfnamefont {Z.}~\bibnamefont
  {Bern}}, \bibinfo {author} {\bibfnamefont {J.~J.}\ \bibnamefont {Carrasco}},
  \bibinfo {author} {\bibfnamefont {L.~J.}\ \bibnamefont {Dixon}}, \bibinfo
  {author} {\bibfnamefont {H.}~\bibnamefont {Johansson}}, \ and\ \bibinfo
  {author} {\bibfnamefont {R.}~\bibnamefont {Roiban}},\ }\href {\doibase
  10.1103/PhysRevLett.103.081301} {\bibfield  {journal} {\bibinfo  {journal}
  {Phys. Rev. Lett.}\ }\textbf {\bibinfo {volume} {103}},\ \bibinfo {pages}
  {081301} (\bibinfo {year} {2009})},\ \Eprint {http://arxiv.org/abs/0905.2326}
  {arXiv:0905.2326 [hep-th]} \BibitemShut {NoStop}%
\bibitem [{\citenamefont {Bern}\ \emph {et~al.}(2012)\citenamefont {Bern},
  \citenamefont {Carrasco}, \citenamefont {Dixon}, \citenamefont {Johansson},\
  and\ \citenamefont {Roiban}}]{Bern:2012uf}%
  \BibitemOpen
  \bibfield  {author} {\bibinfo {author} {\bibfnamefont {Z.}~\bibnamefont
  {Bern}}, \bibinfo {author} {\bibfnamefont {J.~J.~M.}\ \bibnamefont
  {Carrasco}}, \bibinfo {author} {\bibfnamefont {L.~J.}\ \bibnamefont {Dixon}},
  \bibinfo {author} {\bibfnamefont {H.}~\bibnamefont {Johansson}}, \ and\
  \bibinfo {author} {\bibfnamefont {R.}~\bibnamefont {Roiban}},\ }\href
  {\doibase 10.1103/PhysRevD.85.105014} {\bibfield  {journal} {\bibinfo
  {journal} {Phys. Rev. D}\ }\textbf {\bibinfo {volume} {85}},\ \bibinfo
  {pages} {105014} (\bibinfo {year} {2012})},\ \Eprint
  {http://arxiv.org/abs/1201.5366} {arXiv:1201.5366 [hep-th]} \BibitemShut
  {NoStop}%
\bibitem [{\citenamefont {Bern}\ \emph
  {et~al.}(2017{\natexlab{a}})\citenamefont {Bern}, \citenamefont {Carrasco},
  \citenamefont {Chen}, \citenamefont {Johansson},\ and\ \citenamefont
  {Roiban}}]{Bern:2017yxu}%
  \BibitemOpen
  \bibfield  {author} {\bibinfo {author} {\bibfnamefont {Z.}~\bibnamefont
  {Bern}}, \bibinfo {author} {\bibfnamefont {J.~J.}\ \bibnamefont {Carrasco}},
  \bibinfo {author} {\bibfnamefont {W.-M.}\ \bibnamefont {Chen}}, \bibinfo
  {author} {\bibfnamefont {H.}~\bibnamefont {Johansson}}, \ and\ \bibinfo
  {author} {\bibfnamefont {R.}~\bibnamefont {Roiban}},\ }\href {\doibase
  10.1103/PhysRevLett.118.181602} {\bibfield  {journal} {\bibinfo  {journal}
  {Phys. Rev. Lett.}\ }\textbf {\bibinfo {volume} {118}},\ \bibinfo {pages}
  {181602} (\bibinfo {year} {2017}{\natexlab{a}})},\ \Eprint
  {http://arxiv.org/abs/1701.02519} {arXiv:1701.02519 [hep-th]} \BibitemShut
  {NoStop}%
\bibitem [{\citenamefont {Bern}\ \emph
  {et~al.}(2017{\natexlab{b}})\citenamefont {Bern}, \citenamefont {Carrasco},
  \citenamefont {Chen}, \citenamefont {Johansson}, \citenamefont {Roiban},\
  and\ \citenamefont {Zeng}}]{Bern:2017ucb}%
  \BibitemOpen
  \bibfield  {author} {\bibinfo {author} {\bibfnamefont {Z.}~\bibnamefont
  {Bern}}, \bibinfo {author} {\bibfnamefont {J.~J.~M.}\ \bibnamefont
  {Carrasco}}, \bibinfo {author} {\bibfnamefont {W.-M.}\ \bibnamefont {Chen}},
  \bibinfo {author} {\bibfnamefont {H.}~\bibnamefont {Johansson}}, \bibinfo
  {author} {\bibfnamefont {R.}~\bibnamefont {Roiban}}, \ and\ \bibinfo {author}
  {\bibfnamefont {M.}~\bibnamefont {Zeng}},\ }\href {\doibase
  10.1103/PhysRevD.96.126012} {\bibfield  {journal} {\bibinfo  {journal} {Phys.
  Rev. D}\ }\textbf {\bibinfo {volume} {96}},\ \bibinfo {pages} {126012}
  (\bibinfo {year} {2017}{\natexlab{b}})},\ \Eprint
  {http://arxiv.org/abs/1708.06807} {arXiv:1708.06807 [hep-th]} \BibitemShut
  {NoStop}%
\bibitem [{\citenamefont {Bern}\ \emph {et~al.}(2018)\citenamefont {Bern},
  \citenamefont {Carrasco}, \citenamefont {Chen}, \citenamefont {Edison},
  \citenamefont {Johansson}, \citenamefont {Parra-Martinez}, \citenamefont
  {Roiban},\ and\ \citenamefont {Zeng}}]{Bern:2018jmv}%
  \BibitemOpen
  \bibfield  {author} {\bibinfo {author} {\bibfnamefont {Z.}~\bibnamefont
  {Bern}}, \bibinfo {author} {\bibfnamefont {J.~J.}\ \bibnamefont {Carrasco}},
  \bibinfo {author} {\bibfnamefont {W.-M.}\ \bibnamefont {Chen}}, \bibinfo
  {author} {\bibfnamefont {A.}~\bibnamefont {Edison}}, \bibinfo {author}
  {\bibfnamefont {H.}~\bibnamefont {Johansson}}, \bibinfo {author}
  {\bibfnamefont {J.}~\bibnamefont {Parra-Martinez}}, \bibinfo {author}
  {\bibfnamefont {R.}~\bibnamefont {Roiban}}, \ and\ \bibinfo {author}
  {\bibfnamefont {M.}~\bibnamefont {Zeng}},\ }\href {\doibase
  10.1103/PhysRevD.98.086021} {\bibfield  {journal} {\bibinfo  {journal} {Phys.
  Rev. D}\ }\textbf {\bibinfo {volume} {98}},\ \bibinfo {pages} {086021}
  (\bibinfo {year} {2018})},\ \Eprint {http://arxiv.org/abs/1804.09311}
  {arXiv:1804.09311 [hep-th]} \BibitemShut {NoStop}%
\bibitem [{\citenamefont {Carrasco}\ and\ \citenamefont
  {Johansson}(2011)}]{Carrasco:2011hw}%
  \BibitemOpen
  \bibfield  {author} {\bibinfo {author} {\bibfnamefont {J.~J.~M.}\
  \bibnamefont {Carrasco}}\ and\ \bibinfo {author} {\bibfnamefont
  {H.}~\bibnamefont {Johansson}},\ }\href {\doibase
  10.1088/1751-8113/44/45/454004} {\bibfield  {journal} {\bibinfo  {journal}
  {J. Phys. A}\ }\textbf {\bibinfo {volume} {44}},\ \bibinfo {pages} {454004}
  (\bibinfo {year} {2011})},\ \Eprint {http://arxiv.org/abs/1103.3298}
  {arXiv:1103.3298 [hep-th]} \BibitemShut {NoStop}%
\bibitem [{\citenamefont {'t~Hooft}(1974)}]{tHooft:1973alw}%
  \BibitemOpen
  \bibfield  {author} {\bibinfo {author} {\bibfnamefont {G.}~\bibnamefont
  {'t~Hooft}},\ }\href {\doibase 10.1016/0550-3213(74)90154-0} {\bibfield
  {journal} {\bibinfo  {journal} {Nucl. Phys. B}\ }\textbf {\bibinfo {volume}
  {72}},\ \bibinfo {pages} {461} (\bibinfo {year} {1974})}\BibitemShut
  {NoStop}%
\bibitem [{\citenamefont {Bern}\ \emph
  {et~al.}(2007{\natexlab{b}})\citenamefont {Bern}, \citenamefont {Czakon},
  \citenamefont {Dixon}, \citenamefont {Kosower},\ and\ \citenamefont
  {Smirnov}}]{Bern:2006ew}%
  \BibitemOpen
  \bibfield  {author} {\bibinfo {author} {\bibfnamefont {Z.}~\bibnamefont
  {Bern}}, \bibinfo {author} {\bibfnamefont {M.}~\bibnamefont {Czakon}},
  \bibinfo {author} {\bibfnamefont {L.~J.}\ \bibnamefont {Dixon}}, \bibinfo
  {author} {\bibfnamefont {D.~A.}\ \bibnamefont {Kosower}}, \ and\ \bibinfo
  {author} {\bibfnamefont {V.~A.}\ \bibnamefont {Smirnov}},\ }\href {\doibase
  10.1103/PhysRevD.75.085010} {\bibfield  {journal} {\bibinfo  {journal} {Phys.
  Rev. D}\ }\textbf {\bibinfo {volume} {75}},\ \bibinfo {pages} {085010}
  (\bibinfo {year} {2007}{\natexlab{b}})},\ \Eprint
  {http://arxiv.org/abs/hep-th/0610248} {arXiv:hep-th/0610248} \BibitemShut
  {NoStop}%
\bibitem [{\citenamefont {Bern}\ \emph
  {et~al.}(2007{\natexlab{c}})\citenamefont {Bern}, \citenamefont {Carrasco},
  \citenamefont {Johansson},\ and\ \citenamefont {Kosower}}]{Bern:2007ct}%
  \BibitemOpen
  \bibfield  {author} {\bibinfo {author} {\bibfnamefont {Z.}~\bibnamefont
  {Bern}}, \bibinfo {author} {\bibfnamefont {J.~J.~M.}\ \bibnamefont
  {Carrasco}}, \bibinfo {author} {\bibfnamefont {H.}~\bibnamefont {Johansson}},
  \ and\ \bibinfo {author} {\bibfnamefont {D.~A.}\ \bibnamefont {Kosower}},\
  }\href {\doibase 10.1103/PhysRevD.76.125020} {\bibfield  {journal} {\bibinfo
  {journal} {Phys. Rev. D}\ }\textbf {\bibinfo {volume} {76}},\ \bibinfo
  {pages} {125020} (\bibinfo {year} {2007}{\natexlab{c}})},\ \Eprint
  {http://arxiv.org/abs/0705.1864} {arXiv:0705.1864 [hep-th]} \BibitemShut
  {NoStop}%
\bibitem [{\citenamefont {Arkani-Hamed}\ \emph {et~al.}(2011)\citenamefont
  {Arkani-Hamed}, \citenamefont {Bourjaily}, \citenamefont {Cachazo},
  \citenamefont {Caron-Huot},\ and\ \citenamefont
  {Trnka}}]{Arkani-Hamed:2010zjl}%
  \BibitemOpen
  \bibfield  {author} {\bibinfo {author} {\bibfnamefont {N.}~\bibnamefont
  {Arkani-Hamed}}, \bibinfo {author} {\bibfnamefont {J.~L.}\ \bibnamefont
  {Bourjaily}}, \bibinfo {author} {\bibfnamefont {F.}~\bibnamefont {Cachazo}},
  \bibinfo {author} {\bibfnamefont {S.}~\bibnamefont {Caron-Huot}}, \ and\
  \bibinfo {author} {\bibfnamefont {J.}~\bibnamefont {Trnka}},\ }\href
  {\doibase 10.1007/JHEP01(2011)041} {\bibfield  {journal} {\bibinfo  {journal}
  {JHEP}\ }\textbf {\bibinfo {volume} {01}},\ \bibinfo {pages} {041} (\bibinfo
  {year} {2011})},\ \Eprint {http://arxiv.org/abs/1008.2958} {arXiv:1008.2958
  [hep-th]} \BibitemShut {NoStop}%
\bibitem [{\citenamefont {Bern}\ \emph {et~al.}(2013)\citenamefont {Bern},
  \citenamefont {Carrasco}, \citenamefont {Dixon}, \citenamefont {Douglas},
  \citenamefont {von Hippel},\ and\ \citenamefont {Johansson}}]{Bern:2012di}%
  \BibitemOpen
  \bibfield  {author} {\bibinfo {author} {\bibfnamefont {Z.}~\bibnamefont
  {Bern}}, \bibinfo {author} {\bibfnamefont {J.~J.}\ \bibnamefont {Carrasco}},
  \bibinfo {author} {\bibfnamefont {L.~J.}\ \bibnamefont {Dixon}}, \bibinfo
  {author} {\bibfnamefont {M.~R.}\ \bibnamefont {Douglas}}, \bibinfo {author}
  {\bibfnamefont {M.}~\bibnamefont {von Hippel}}, \ and\ \bibinfo {author}
  {\bibfnamefont {H.}~\bibnamefont {Johansson}},\ }\href {\doibase
  10.1103/PhysRevD.87.025018} {\bibfield  {journal} {\bibinfo  {journal} {Phys.
  Rev. D}\ }\textbf {\bibinfo {volume} {87}},\ \bibinfo {pages} {025018}
  (\bibinfo {year} {2013})},\ \Eprint {http://arxiv.org/abs/1210.7709}
  {arXiv:1210.7709 [hep-th]} \BibitemShut {NoStop}%
\bibitem [{\citenamefont {Bourjaily}\ \emph {et~al.}(2016)\citenamefont
  {Bourjaily}, \citenamefont {Heslop},\ and\ \citenamefont
  {Tran}}]{Bourjaily:2016evz}%
  \BibitemOpen
  \bibfield  {author} {\bibinfo {author} {\bibfnamefont {J.~L.}\ \bibnamefont
  {Bourjaily}}, \bibinfo {author} {\bibfnamefont {P.}~\bibnamefont {Heslop}}, \
  and\ \bibinfo {author} {\bibfnamefont {V.-V.}\ \bibnamefont {Tran}},\ }\href
  {\doibase 10.1007/JHEP11(2016)125} {\bibfield  {journal} {\bibinfo  {journal}
  {JHEP}\ }\textbf {\bibinfo {volume} {11}},\ \bibinfo {pages} {125} (\bibinfo
  {year} {2016})},\ \Eprint {http://arxiv.org/abs/1609.00007} {arXiv:1609.00007
  [hep-th]} \BibitemShut {NoStop}%
\bibitem [{\citenamefont {Drummond}\ \emph {et~al.}(2007)\citenamefont
  {Drummond}, \citenamefont {Henn}, \citenamefont {Smirnov},\ and\
  \citenamefont {Sokatchev}}]{Drummond:2006rz}%
  \BibitemOpen
  \bibfield  {author} {\bibinfo {author} {\bibfnamefont {J.~M.}\ \bibnamefont
  {Drummond}}, \bibinfo {author} {\bibfnamefont {J.}~\bibnamefont {Henn}},
  \bibinfo {author} {\bibfnamefont {V.~A.}\ \bibnamefont {Smirnov}}, \ and\
  \bibinfo {author} {\bibfnamefont {E.}~\bibnamefont {Sokatchev}},\ }\href
  {\doibase 10.1088/1126-6708/2007/01/064} {\bibfield  {journal} {\bibinfo
  {journal} {JHEP}\ }\textbf {\bibinfo {volume} {01}},\ \bibinfo {pages} {064}
  (\bibinfo {year} {2007})},\ \Eprint {http://arxiv.org/abs/hep-th/0607160}
  {arXiv:hep-th/0607160} \BibitemShut {NoStop}%
\bibitem [{\citenamefont {Drummond}\ \emph {et~al.}(2010)\citenamefont
  {Drummond}, \citenamefont {Henn}, \citenamefont {Korchemsky},\ and\
  \citenamefont {Sokatchev}}]{Drummond:2008vq}%
  \BibitemOpen
  \bibfield  {author} {\bibinfo {author} {\bibfnamefont {J.~M.}\ \bibnamefont
  {Drummond}}, \bibinfo {author} {\bibfnamefont {J.}~\bibnamefont {Henn}},
  \bibinfo {author} {\bibfnamefont {G.~P.}\ \bibnamefont {Korchemsky}}, \ and\
  \bibinfo {author} {\bibfnamefont {E.}~\bibnamefont {Sokatchev}},\ }\href
  {\doibase 10.1016/j.nuclphysb.2009.11.022} {\bibfield  {journal} {\bibinfo
  {journal} {Nucl. Phys. B}\ }\textbf {\bibinfo {volume} {828}},\ \bibinfo
  {pages} {317} (\bibinfo {year} {2010})},\ \Eprint
  {http://arxiv.org/abs/0807.1095} {arXiv:0807.1095 [hep-th]} \BibitemShut
  {NoStop}%
\bibitem [{\citenamefont {Brandhuber}\ \emph {et~al.}(2008)\citenamefont
  {Brandhuber}, \citenamefont {Heslop},\ and\ \citenamefont
  {Travaglini}}]{Brandhuber:2008pf}%
  \BibitemOpen
  \bibfield  {author} {\bibinfo {author} {\bibfnamefont {A.}~\bibnamefont
  {Brandhuber}}, \bibinfo {author} {\bibfnamefont {P.}~\bibnamefont {Heslop}},
  \ and\ \bibinfo {author} {\bibfnamefont {G.}~\bibnamefont {Travaglini}},\
  }\href {\doibase 10.1103/PhysRevD.78.125005} {\bibfield  {journal} {\bibinfo
  {journal} {Phys. Rev. D}\ }\textbf {\bibinfo {volume} {78}},\ \bibinfo
  {pages} {125005} (\bibinfo {year} {2008})},\ \Eprint
  {http://arxiv.org/abs/0807.4097} {arXiv:0807.4097 [hep-th]} \BibitemShut
  {NoStop}%
\bibitem [{\citenamefont {Drummond}\ \emph {et~al.}(2009)\citenamefont
  {Drummond}, \citenamefont {Henn},\ and\ \citenamefont
  {Plefka}}]{Drummond:2009fd}%
  \BibitemOpen
  \bibfield  {author} {\bibinfo {author} {\bibfnamefont {J.~M.}\ \bibnamefont
  {Drummond}}, \bibinfo {author} {\bibfnamefont {J.~M.}\ \bibnamefont {Henn}},
  \ and\ \bibinfo {author} {\bibfnamefont {J.}~\bibnamefont {Plefka}},\ }\href
  {\doibase 10.1088/1126-6708/2009/05/046} {\bibfield  {journal} {\bibinfo
  {journal} {JHEP}\ }\textbf {\bibinfo {volume} {05}},\ \bibinfo {pages} {046}
  (\bibinfo {year} {2009})},\ \Eprint {http://arxiv.org/abs/0902.2987}
  {arXiv:0902.2987 [hep-th]} \BibitemShut {NoStop}%
\bibitem [{\citenamefont {Arkani-Hamed}\ \emph {et~al.}(2016)\citenamefont
  {Arkani-Hamed}, \citenamefont {Bourjaily}, \citenamefont {Cachazo},
  \citenamefont {Goncharov}, \citenamefont {Postnikov},\ and\ \citenamefont
  {Trnka}}]{Arkani-Hamed:2012zlh}%
  \BibitemOpen
  \bibfield  {author} {\bibinfo {author} {\bibfnamefont {N.}~\bibnamefont
  {Arkani-Hamed}}, \bibinfo {author} {\bibfnamefont {J.~L.}\ \bibnamefont
  {Bourjaily}}, \bibinfo {author} {\bibfnamefont {F.}~\bibnamefont {Cachazo}},
  \bibinfo {author} {\bibfnamefont {A.~B.}\ \bibnamefont {Goncharov}}, \bibinfo
  {author} {\bibfnamefont {A.}~\bibnamefont {Postnikov}}, \ and\ \bibinfo
  {author} {\bibfnamefont {J.}~\bibnamefont {Trnka}},\ }\href {\doibase
  10.1017/CBO9781316091548} {\emph {\bibinfo {title} {{Grassmannian Geometry of
  Scattering Amplitudes}}}}\ (\bibinfo  {publisher} {Cambridge University
  Press},\ \bibinfo {year} {2016})\ \Eprint {http://arxiv.org/abs/1212.5605}
  {arXiv:1212.5605 [hep-th]} \BibitemShut {NoStop}%
\bibitem [{\citenamefont {Arkani-Hamed}\ and\ \citenamefont
  {Trnka}(2014)}]{Arkani-Hamed:2013jha}%
  \BibitemOpen
  \bibfield  {author} {\bibinfo {author} {\bibfnamefont {N.}~\bibnamefont
  {Arkani-Hamed}}\ and\ \bibinfo {author} {\bibfnamefont {J.}~\bibnamefont
  {Trnka}},\ }\href {\doibase 10.1007/JHEP10(2014)030} {\bibfield  {journal}
  {\bibinfo  {journal} {JHEP}\ }\textbf {\bibinfo {volume} {10}},\ \bibinfo
  {pages} {030} (\bibinfo {year} {2014})},\ \Eprint
  {http://arxiv.org/abs/1312.2007} {arXiv:1312.2007 [hep-th]} \BibitemShut
  {NoStop}%
\bibitem [{\citenamefont {Arkani-Hamed}\ \emph {et~al.}(2015)\citenamefont
  {Arkani-Hamed}, \citenamefont {Hodges},\ and\ \citenamefont
  {Trnka}}]{Arkani-Hamed:2014dca}%
  \BibitemOpen
  \bibfield  {author} {\bibinfo {author} {\bibfnamefont {N.}~\bibnamefont
  {Arkani-Hamed}}, \bibinfo {author} {\bibfnamefont {A.}~\bibnamefont
  {Hodges}}, \ and\ \bibinfo {author} {\bibfnamefont {J.}~\bibnamefont
  {Trnka}},\ }\href {\doibase 10.1007/JHEP08(2015)030} {\bibfield  {journal}
  {\bibinfo  {journal} {JHEP}\ }\textbf {\bibinfo {volume} {08}},\ \bibinfo
  {pages} {030} (\bibinfo {year} {2015})},\ \Eprint
  {http://arxiv.org/abs/1412.8478} {arXiv:1412.8478 [hep-th]} \BibitemShut
  {NoStop}%
\bibitem [{\citenamefont {Arkani-Hamed}\ \emph {et~al.}(2019)\citenamefont
  {Arkani-Hamed}, \citenamefont {Langer}, \citenamefont {Yelleshpur~Srikant},\
  and\ \citenamefont {Trnka}}]{Arkani-Hamed:2018rsk}%
  \BibitemOpen
  \bibfield  {author} {\bibinfo {author} {\bibfnamefont {N.}~\bibnamefont
  {Arkani-Hamed}}, \bibinfo {author} {\bibfnamefont {C.}~\bibnamefont
  {Langer}}, \bibinfo {author} {\bibfnamefont {A.}~\bibnamefont
  {Yelleshpur~Srikant}}, \ and\ \bibinfo {author} {\bibfnamefont
  {J.}~\bibnamefont {Trnka}},\ }\href {\doibase 10.1103/PhysRevLett.122.051601}
  {\bibfield  {journal} {\bibinfo  {journal} {Phys. Rev. Lett.}\ }\textbf
  {\bibinfo {volume} {122}},\ \bibinfo {pages} {051601} (\bibinfo {year}
  {2019})},\ \Eprint {http://arxiv.org/abs/1810.08208} {arXiv:1810.08208
  [hep-th]} \BibitemShut {NoStop}%
\bibitem [{\citenamefont {Langer}\ and\ \citenamefont
  {Yelleshpur~Srikant}(2019)}]{Langer:2019iuo}%
  \BibitemOpen
  \bibfield  {author} {\bibinfo {author} {\bibfnamefont {C.}~\bibnamefont
  {Langer}}\ and\ \bibinfo {author} {\bibfnamefont {A.}~\bibnamefont
  {Yelleshpur~Srikant}},\ }\href {\doibase 10.1007/JHEP04(2019)105} {\bibfield
  {journal} {\bibinfo  {journal} {JHEP}\ }\textbf {\bibinfo {volume} {04}},\
  \bibinfo {pages} {105} (\bibinfo {year} {2019})},\ \Eprint
  {http://arxiv.org/abs/1902.05951} {arXiv:1902.05951 [hep-th]} \BibitemShut
  {NoStop}%
\bibitem [{\citenamefont {Bern}\ \emph
  {et~al.}(2010{\natexlab{b}})\citenamefont {Bern}, \citenamefont {Carrasco},
  \citenamefont {Dixon}, \citenamefont {Johansson},\ and\ \citenamefont
  {Roiban}}]{Bern:2010tq}%
  \BibitemOpen
  \bibfield  {author} {\bibinfo {author} {\bibfnamefont {Z.}~\bibnamefont
  {Bern}}, \bibinfo {author} {\bibfnamefont {J.~J.~M.}\ \bibnamefont
  {Carrasco}}, \bibinfo {author} {\bibfnamefont {L.~J.}\ \bibnamefont {Dixon}},
  \bibinfo {author} {\bibfnamefont {H.}~\bibnamefont {Johansson}}, \ and\
  \bibinfo {author} {\bibfnamefont {R.}~\bibnamefont {Roiban}},\ }\href
  {\doibase 10.1103/PhysRevD.82.125040} {\bibfield  {journal} {\bibinfo
  {journal} {Phys. Rev. D}\ }\textbf {\bibinfo {volume} {82}},\ \bibinfo
  {pages} {125040} (\bibinfo {year} {2010}{\natexlab{b}})},\ \Eprint
  {http://arxiv.org/abs/1008.3327} {arXiv:1008.3327 [hep-th]} \BibitemShut
  {NoStop}%
\bibitem [{\citenamefont {Bourjaily}\ \emph {et~al.}(2012)\citenamefont
  {Bourjaily}, \citenamefont {DiRe}, \citenamefont {Shaikh}, \citenamefont
  {Spradlin},\ and\ \citenamefont {Volovich}}]{Bourjaily:2011hi}%
  \BibitemOpen
  \bibfield  {author} {\bibinfo {author} {\bibfnamefont {J.~L.}\ \bibnamefont
  {Bourjaily}}, \bibinfo {author} {\bibfnamefont {A.}~\bibnamefont {DiRe}},
  \bibinfo {author} {\bibfnamefont {A.}~\bibnamefont {Shaikh}}, \bibinfo
  {author} {\bibfnamefont {M.}~\bibnamefont {Spradlin}}, \ and\ \bibinfo
  {author} {\bibfnamefont {A.}~\bibnamefont {Volovich}},\ }\href {\doibase
  10.1007/JHEP03(2012)032} {\bibfield  {journal} {\bibinfo  {journal} {JHEP}\
  }\textbf {\bibinfo {volume} {03}},\ \bibinfo {pages} {032} (\bibinfo {year}
  {2012})},\ \Eprint {http://arxiv.org/abs/1112.6432} {arXiv:1112.6432
  [hep-th]} \BibitemShut {NoStop}%
\bibitem [{\citenamefont {Eden}\ \emph {et~al.}(2012)\citenamefont {Eden},
  \citenamefont {Heslop}, \citenamefont {Korchemsky},\ and\ \citenamefont
  {Sokatchev}}]{Eden:2012tu}%
  \BibitemOpen
  \bibfield  {author} {\bibinfo {author} {\bibfnamefont {B.}~\bibnamefont
  {Eden}}, \bibinfo {author} {\bibfnamefont {P.}~\bibnamefont {Heslop}},
  \bibinfo {author} {\bibfnamefont {G.~P.}\ \bibnamefont {Korchemsky}}, \ and\
  \bibinfo {author} {\bibfnamefont {E.}~\bibnamefont {Sokatchev}},\ }\href
  {\doibase 10.1016/j.nuclphysb.2012.04.013} {\bibfield  {journal} {\bibinfo
  {journal} {Nucl. Phys. B}\ }\textbf {\bibinfo {volume} {862}},\ \bibinfo
  {pages} {450} (\bibinfo {year} {2012})},\ \Eprint
  {http://arxiv.org/abs/1201.5329} {arXiv:1201.5329 [hep-th]} \BibitemShut
  {NoStop}%
\bibitem [{\citenamefont {Bern}\ \emph {et~al.}(1997)\citenamefont {Bern},
  \citenamefont {Rozowsky},\ and\ \citenamefont {Yan}}]{Bern:1997nh}%
  \BibitemOpen
  \bibfield  {author} {\bibinfo {author} {\bibfnamefont {Z.}~\bibnamefont
  {Bern}}, \bibinfo {author} {\bibfnamefont {J.~S.}\ \bibnamefont {Rozowsky}},
  \ and\ \bibinfo {author} {\bibfnamefont {B.}~\bibnamefont {Yan}},\ }\href
  {\doibase 10.1016/S0370-2693(97)00413-9} {\bibfield  {journal} {\bibinfo
  {journal} {Phys. Lett. B}\ }\textbf {\bibinfo {volume} {401}},\ \bibinfo
  {pages} {273} (\bibinfo {year} {1997})},\ \Eprint
  {http://arxiv.org/abs/hep-ph/9702424} {arXiv:hep-ph/9702424} \BibitemShut
  {NoStop}%
\bibitem [{\citenamefont {Bern}\ \emph
  {et~al.}(1998{\natexlab{a}})\citenamefont {Bern}, \citenamefont {Dixon},
  \citenamefont {Dunbar}, \citenamefont {Perelstein},\ and\ \citenamefont
  {Rozowsky}}]{Bern:1998ug}%
  \BibitemOpen
  \bibfield  {author} {\bibinfo {author} {\bibfnamefont {Z.}~\bibnamefont
  {Bern}}, \bibinfo {author} {\bibfnamefont {L.~J.}\ \bibnamefont {Dixon}},
  \bibinfo {author} {\bibfnamefont {D.~C.}\ \bibnamefont {Dunbar}}, \bibinfo
  {author} {\bibfnamefont {M.}~\bibnamefont {Perelstein}}, \ and\ \bibinfo
  {author} {\bibfnamefont {J.~S.}\ \bibnamefont {Rozowsky}},\ }\href {\doibase
  10.1016/S0550-3213(98)00420-9} {\bibfield  {journal} {\bibinfo  {journal}
  {Nucl. Phys. B}\ }\textbf {\bibinfo {volume} {530}},\ \bibinfo {pages} {401}
  (\bibinfo {year} {1998}{\natexlab{a}})},\ \Eprint
  {http://arxiv.org/abs/hep-th/9802162} {arXiv:hep-th/9802162} \BibitemShut
  {NoStop}%
\bibitem [{\citenamefont {Weinberg}(1964)}]{Weinberg:1964ew}%
  \BibitemOpen
  \bibfield  {author} {\bibinfo {author} {\bibfnamefont {S.}~\bibnamefont
  {Weinberg}},\ }\href {\doibase 10.1103/PhysRev.135.B1049} {\bibfield
  {journal} {\bibinfo  {journal} {Phys. Rev.}\ }\textbf {\bibinfo {volume}
  {135}},\ \bibinfo {pages} {B1049} (\bibinfo {year} {1964})}\BibitemShut
  {NoStop}%
\bibitem [{\citenamefont {Weinberg}(1965)}]{Weinberg:1965nx}%
  \BibitemOpen
  \bibfield  {author} {\bibinfo {author} {\bibfnamefont {S.}~\bibnamefont
  {Weinberg}},\ }\href {\doibase 10.1103/PhysRev.140.B516} {\bibfield
  {journal} {\bibinfo  {journal} {Phys. Rev.}\ }\textbf {\bibinfo {volume}
  {140}},\ \bibinfo {pages} {B516} (\bibinfo {year} {1965})}\BibitemShut
  {NoStop}%
\bibitem [{\citenamefont {Low}(1954)}]{Low:1954kd}%
  \BibitemOpen
  \bibfield  {author} {\bibinfo {author} {\bibfnamefont {F.~E.}\ \bibnamefont
  {Low}},\ }\href {\doibase 10.1103/PhysRev.96.1428} {\bibfield  {journal}
  {\bibinfo  {journal} {Phys. Rev.}\ }\textbf {\bibinfo {volume} {96}},\
  \bibinfo {pages} {1428} (\bibinfo {year} {1954})}\BibitemShut {NoStop}%
\bibitem [{\citenamefont {Gell-Mann}\ and\ \citenamefont
  {Goldberger}(1954)}]{Gell-Mann:1954wra}%
  \BibitemOpen
  \bibfield  {author} {\bibinfo {author} {\bibfnamefont {M.}~\bibnamefont
  {Gell-Mann}}\ and\ \bibinfo {author} {\bibfnamefont {M.~L.}\ \bibnamefont
  {Goldberger}},\ }\href {\doibase 10.1103/PhysRev.96.1433} {\bibfield
  {journal} {\bibinfo  {journal} {Phys. Rev.}\ }\textbf {\bibinfo {volume}
  {96}},\ \bibinfo {pages} {1433} (\bibinfo {year} {1954})}\BibitemShut
  {NoStop}%
\bibitem [{\citenamefont {Gross}\ and\ \citenamefont
  {Jackiw}(1968)}]{Gross:1968in}%
  \BibitemOpen
  \bibfield  {author} {\bibinfo {author} {\bibfnamefont {D.~J.}\ \bibnamefont
  {Gross}}\ and\ \bibinfo {author} {\bibfnamefont {R.}~\bibnamefont {Jackiw}},\
  }\href {\doibase 10.1103/PhysRev.166.1287} {\bibfield  {journal} {\bibinfo
  {journal} {Phys. Rev.}\ }\textbf {\bibinfo {volume} {166}},\ \bibinfo {pages}
  {1287} (\bibinfo {year} {1968})}\BibitemShut {NoStop}%
\bibitem [{\citenamefont {Laenen}\ \emph {et~al.}(2009)\citenamefont {Laenen},
  \citenamefont {Stavenga},\ and\ \citenamefont {White}}]{Laenen:2008gt}%
  \BibitemOpen
  \bibfield  {author} {\bibinfo {author} {\bibfnamefont {E.}~\bibnamefont
  {Laenen}}, \bibinfo {author} {\bibfnamefont {G.}~\bibnamefont {Stavenga}}, \
  and\ \bibinfo {author} {\bibfnamefont {C.~D.}\ \bibnamefont {White}},\ }\href
  {\doibase 10.1088/1126-6708/2009/03/054} {\bibfield  {journal} {\bibinfo
  {journal} {JHEP}\ }\textbf {\bibinfo {volume} {03}},\ \bibinfo {pages} {054}
  (\bibinfo {year} {2009})},\ \Eprint {http://arxiv.org/abs/0811.2067}
  {arXiv:0811.2067 [hep-ph]} \BibitemShut {NoStop}%
\bibitem [{\citenamefont {White}(2011)}]{White:2011yy}%
  \BibitemOpen
  \bibfield  {author} {\bibinfo {author} {\bibfnamefont {C.~D.}\ \bibnamefont
  {White}},\ }\href {\doibase 10.1007/JHEP05(2011)060} {\bibfield  {journal}
  {\bibinfo  {journal} {JHEP}\ }\textbf {\bibinfo {volume} {05}},\ \bibinfo
  {pages} {060} (\bibinfo {year} {2011})},\ \Eprint
  {http://arxiv.org/abs/1103.2981} {arXiv:1103.2981 [hep-th]} \BibitemShut
  {NoStop}%
\bibitem [{\citenamefont {Bern}\ \emph
  {et~al.}(1998{\natexlab{b}})\citenamefont {Bern}, \citenamefont {Dixon},\
  and\ \citenamefont {Kosower}}]{Bern:1997sc}%
  \BibitemOpen
  \bibfield  {author} {\bibinfo {author} {\bibfnamefont {Z.}~\bibnamefont
  {Bern}}, \bibinfo {author} {\bibfnamefont {L.~J.}\ \bibnamefont {Dixon}}, \
  and\ \bibinfo {author} {\bibfnamefont {D.~A.}\ \bibnamefont {Kosower}},\
  }\href {\doibase 10.1016/S0550-3213(97)00703-7} {\bibfield  {journal}
  {\bibinfo  {journal} {Nucl. Phys. B}\ }\textbf {\bibinfo {volume} {513}},\
  \bibinfo {pages} {3} (\bibinfo {year} {1998}{\natexlab{b}})},\ \Eprint
  {http://arxiv.org/abs/hep-ph/9708239} {arXiv:hep-ph/9708239} \BibitemShut
  {NoStop}%
\bibitem [{\citenamefont {Bern}\ \emph {et~al.}(2005)\citenamefont {Bern},
  \citenamefont {Del~Duca}, \citenamefont {Dixon},\ and\ \citenamefont
  {Kosower}}]{Bern:2004ky}%
  \BibitemOpen
  \bibfield  {author} {\bibinfo {author} {\bibfnamefont {Z.}~\bibnamefont
  {Bern}}, \bibinfo {author} {\bibfnamefont {V.}~\bibnamefont {Del~Duca}},
  \bibinfo {author} {\bibfnamefont {L.~J.}\ \bibnamefont {Dixon}}, \ and\
  \bibinfo {author} {\bibfnamefont {D.~A.}\ \bibnamefont {Kosower}},\ }\href
  {\doibase 10.1103/PhysRevD.71.045006} {\bibfield  {journal} {\bibinfo
  {journal} {Phys. Rev. D}\ }\textbf {\bibinfo {volume} {71}},\ \bibinfo
  {pages} {045006} (\bibinfo {year} {2005})},\ \Eprint
  {http://arxiv.org/abs/hep-th/0410224} {arXiv:hep-th/0410224} \BibitemShut
  {NoStop}%
\bibitem [{\citenamefont {Britto}\ \emph {et~al.}(2005)\citenamefont {Britto},
  \citenamefont {Cachazo},\ and\ \citenamefont {Feng}}]{Britto:2004nc}%
  \BibitemOpen
  \bibfield  {author} {\bibinfo {author} {\bibfnamefont {R.}~\bibnamefont
  {Britto}}, \bibinfo {author} {\bibfnamefont {F.}~\bibnamefont {Cachazo}}, \
  and\ \bibinfo {author} {\bibfnamefont {B.}~\bibnamefont {Feng}},\ }\href
  {\doibase 10.1016/j.nuclphysb.2005.07.014} {\bibfield  {journal} {\bibinfo
  {journal} {Nucl. Phys. B}\ }\textbf {\bibinfo {volume} {725}},\ \bibinfo
  {pages} {275} (\bibinfo {year} {2005})},\ \Eprint
  {http://arxiv.org/abs/hep-th/0412103} {arXiv:hep-th/0412103} \BibitemShut
  {NoStop}%
\bibitem [{\citenamefont {Bussemake}\ \emph {et~al.}(1977)\citenamefont
  {Bussemake}, \citenamefont {Čobeljić}, \citenamefont {Cvetković},\ and\
  \citenamefont {Seidel}}]{BUSSEMAKE1977234}%
  \BibitemOpen
  \bibfield  {author} {\bibinfo {author} {\bibfnamefont {F.}~\bibnamefont
  {Bussemake}}, \bibinfo {author} {\bibfnamefont {S.}~\bibnamefont
  {Čobeljić}}, \bibinfo {author} {\bibfnamefont {D.}~\bibnamefont
  {Cvetković}}, \ and\ \bibinfo {author} {\bibfnamefont {J.}~\bibnamefont
  {Seidel}},\ }\href {\doibase https://doi.org/10.1016/0095-8956(77)90034-X}
  {\bibfield  {journal} {\bibinfo  {journal} {Journal of Combinatorial Theory,
  Series B}\ }\textbf {\bibinfo {volume} {23}},\ \bibinfo {pages} {234}
  (\bibinfo {year} {1977})}\BibitemShut {NoStop}%
\bibitem [{\citenamefont {Bjerrum-Bohr}\ \emph {et~al.}(2006)\citenamefont
  {Bjerrum-Bohr}, \citenamefont {Dunbar}, \citenamefont {Ita}, \citenamefont
  {Perkins},\ and\ \citenamefont {Risager}}]{Bjerrum-Bohr:2006xbk}%
  \BibitemOpen
  \bibfield  {author} {\bibinfo {author} {\bibfnamefont {N.~E.~J.}\
  \bibnamefont {Bjerrum-Bohr}}, \bibinfo {author} {\bibfnamefont {D.~C.}\
  \bibnamefont {Dunbar}}, \bibinfo {author} {\bibfnamefont {H.}~\bibnamefont
  {Ita}}, \bibinfo {author} {\bibfnamefont {W.~B.}\ \bibnamefont {Perkins}}, \
  and\ \bibinfo {author} {\bibfnamefont {K.}~\bibnamefont {Risager}},\ }\href
  {\doibase 10.1088/1126-6708/2006/12/072} {\bibfield  {journal} {\bibinfo
  {journal} {JHEP}\ }\textbf {\bibinfo {volume} {12}},\ \bibinfo {pages} {072}
  (\bibinfo {year} {2006})},\ \Eprint {http://arxiv.org/abs/hep-th/0610043}
  {arXiv:hep-th/0610043} \BibitemShut {NoStop}%
\bibitem [{\citenamefont {Edison}\ and\ \citenamefont
  {Parra-Martinez}()}]{AEJPM}%
  \BibitemOpen
  \bibfield  {author} {\bibinfo {author} {\bibfnamefont {A.}~\bibnamefont
  {Edison}}\ and\ \bibinfo {author} {\bibfnamefont {J.}~\bibnamefont
  {Parra-Martinez}},\ }\href@noop {} {\bibinfo  {journal} {in preparation}\
  }\BibitemShut {NoStop}%
\bibitem [{\citenamefont {Carrasco}\ \emph {et~al.}()\citenamefont {Carrasco},
  \citenamefont {Edison}, \citenamefont {Johansson},\ and\ \citenamefont
  {Verbeek}}]{Carrasco:2022zzz}%
  \BibitemOpen
\bibfield  {journal} {  }\bibfield  {author} {\bibinfo {author} {\bibfnamefont
  {J.~J.~M.}\ \bibnamefont {Carrasco}}, \bibinfo {author} {\bibfnamefont
  {A.}~\bibnamefont {Edison}}, \bibinfo {author} {\bibfnamefont
  {H.}~\bibnamefont {Johansson}}, \ and\ \bibinfo {author} {\bibfnamefont
  {B.}~\bibnamefont {Verbeek}},\ }\href@noop {} {\bibinfo  {journal} {in
  preparation}\ }\BibitemShut {NoStop}%
\bibitem [{\citenamefont {Carrasco}\ \emph {et~al.}(2021)\citenamefont
  {Carrasco}, \citenamefont {Edison},\ and\ \citenamefont
  {Johansson}}]{SixIntZenodo}%
  \BibitemOpen
\bibfield  {journal} {  }\bibfield  {author} {\bibinfo {author} {\bibfnamefont
  {J.~J.~M.}\ \bibnamefont {Carrasco}}, \bibinfo {author} {\bibfnamefont
  {A.}~\bibnamefont {Edison}}, \ and\ \bibinfo {author} {\bibfnamefont
  {H.}~\bibnamefont {Johansson}},\ }\href {\doibase 10.5281/zenodo.5765781}
  {\enquote {\bibinfo {title} {{Four-point six-loop super-Yang-Mills
  integrand}},}\ }\bibinfo {howpublished} {Hosted by Zenodo:
  \texttt{10.5281/zenodo.5765781}} (\bibinfo {year} {2021})\BibitemShut
  {NoStop}%
\bibitem [{\citenamefont {Golden}\ and\ \citenamefont
  {Spradlin}(2012)}]{Golden:2012hi}%
  \BibitemOpen
  \bibfield  {author} {\bibinfo {author} {\bibfnamefont {J.}~\bibnamefont
  {Golden}}\ and\ \bibinfo {author} {\bibfnamefont {M.}~\bibnamefont
  {Spradlin}},\ }\href {\doibase 10.1007/JHEP05(2012)027} {\bibfield  {journal}
  {\bibinfo  {journal} {JHEP}\ }\textbf {\bibinfo {volume} {05}},\ \bibinfo
  {pages} {027} (\bibinfo {year} {2012})},\ \Eprint
  {http://arxiv.org/abs/1203.1915} {arXiv:1203.1915 [hep-th]} \BibitemShut
  {NoStop}%
\bibitem [{\citenamefont {Bourjaily}\ and\ \citenamefont
  {Trnka}(2015)}]{Bourjaily:2015jna}%
  \BibitemOpen
  \bibfield  {author} {\bibinfo {author} {\bibfnamefont {J.~L.}\ \bibnamefont
  {Bourjaily}}\ and\ \bibinfo {author} {\bibfnamefont {J.}~\bibnamefont
  {Trnka}},\ }\href {\doibase 10.1007/JHEP08(2015)119} {\bibfield  {journal}
  {\bibinfo  {journal} {JHEP}\ }\textbf {\bibinfo {volume} {08}},\ \bibinfo
  {pages} {119} (\bibinfo {year} {2015})},\ \Eprint
  {http://arxiv.org/abs/1505.05886} {arXiv:1505.05886 [hep-th]} \BibitemShut
  {NoStop}%
\bibitem [{\citenamefont {Bourjaily}\ \emph {et~al.}(2017)\citenamefont
  {Bourjaily}, \citenamefont {Herrmann},\ and\ \citenamefont
  {Trnka}}]{Bourjaily:2017wjl}%
  \BibitemOpen
  \bibfield  {author} {\bibinfo {author} {\bibfnamefont {J.~L.}\ \bibnamefont
  {Bourjaily}}, \bibinfo {author} {\bibfnamefont {E.}~\bibnamefont {Herrmann}},
  \ and\ \bibinfo {author} {\bibfnamefont {J.}~\bibnamefont {Trnka}},\ }\href
  {\doibase 10.1007/JHEP06(2017)059} {\bibfield  {journal} {\bibinfo  {journal}
  {JHEP}\ }\textbf {\bibinfo {volume} {06}},\ \bibinfo {pages} {059} (\bibinfo
  {year} {2017})},\ \Eprint {http://arxiv.org/abs/1704.05460} {arXiv:1704.05460
  [hep-th]} \BibitemShut {NoStop}%
\bibitem [{\citenamefont {Mafra}\ and\ \citenamefont
  {Schlotterer}(2015)}]{Mafra:2015mja}%
  \BibitemOpen
  \bibfield  {author} {\bibinfo {author} {\bibfnamefont {C.~R.}\ \bibnamefont
  {Mafra}}\ and\ \bibinfo {author} {\bibfnamefont {O.}~\bibnamefont
  {Schlotterer}},\ }\href {\doibase 10.1007/JHEP10(2015)124} {\bibfield
  {journal} {\bibinfo  {journal} {JHEP}\ }\textbf {\bibinfo {volume} {10}},\
  \bibinfo {pages} {124} (\bibinfo {year} {2015})},\ \Eprint
  {http://arxiv.org/abs/1505.02746} {arXiv:1505.02746 [hep-th]} \BibitemShut
  {NoStop}%
\bibitem [{\citenamefont {Bourjaily}\ \emph {et~al.}(2019)\citenamefont
  {Bourjaily}, \citenamefont {Herrmann}, \citenamefont {Langer}, \citenamefont
  {McLeod},\ and\ \citenamefont {Trnka}}]{Bourjaily:2019iqr}%
  \BibitemOpen
  \bibfield  {author} {\bibinfo {author} {\bibfnamefont {J.~L.}\ \bibnamefont
  {Bourjaily}}, \bibinfo {author} {\bibfnamefont {E.}~\bibnamefont {Herrmann}},
  \bibinfo {author} {\bibfnamefont {C.}~\bibnamefont {Langer}}, \bibinfo
  {author} {\bibfnamefont {A.~J.}\ \bibnamefont {McLeod}}, \ and\ \bibinfo
  {author} {\bibfnamefont {J.}~\bibnamefont {Trnka}},\ }\href {\doibase
  10.1007/JHEP12(2019)073} {\bibfield  {journal} {\bibinfo  {journal} {JHEP}\
  }\textbf {\bibinfo {volume} {12}},\ \bibinfo {pages} {073} (\bibinfo {year}
  {2019})},\ \Eprint {http://arxiv.org/abs/1909.09131} {arXiv:1909.09131
  [hep-th]} \BibitemShut {NoStop}%
\bibitem [{\citenamefont {Bourjaily}\ \emph {et~al.}(2020)\citenamefont
  {Bourjaily}, \citenamefont {Herrmann}, \citenamefont {Langer}, \citenamefont
  {McLeod},\ and\ \citenamefont {Trnka}}]{Bourjaily:2019gqu}%
  \BibitemOpen
  \bibfield  {author} {\bibinfo {author} {\bibfnamefont {J.~L.}\ \bibnamefont
  {Bourjaily}}, \bibinfo {author} {\bibfnamefont {E.}~\bibnamefont {Herrmann}},
  \bibinfo {author} {\bibfnamefont {C.}~\bibnamefont {Langer}}, \bibinfo
  {author} {\bibfnamefont {A.~J.}\ \bibnamefont {McLeod}}, \ and\ \bibinfo
  {author} {\bibfnamefont {J.}~\bibnamefont {Trnka}},\ }\href {\doibase
  10.1103/PhysRevLett.124.111603} {\bibfield  {journal} {\bibinfo  {journal}
  {Phys. Rev. Lett.}\ }\textbf {\bibinfo {volume} {124}},\ \bibinfo {pages}
  {111603} (\bibinfo {year} {2020})},\ \Eprint
  {http://arxiv.org/abs/1911.09106} {arXiv:1911.09106 [hep-th]} \BibitemShut
  {NoStop}%
\bibitem [{\citenamefont {Gomez}\ and\ \citenamefont
  {Mafra}(2013)}]{Gomez:2013sla}%
  \BibitemOpen
  \bibfield  {author} {\bibinfo {author} {\bibfnamefont {H.}~\bibnamefont
  {Gomez}}\ and\ \bibinfo {author} {\bibfnamefont {C.~R.}\ \bibnamefont
  {Mafra}},\ }\href {\doibase 10.1007/JHEP10(2013)217} {\bibfield  {journal}
  {\bibinfo  {journal} {JHEP}\ }\textbf {\bibinfo {volume} {10}},\ \bibinfo
  {pages} {217} (\bibinfo {year} {2013})},\ \Eprint
  {http://arxiv.org/abs/1308.6567} {arXiv:1308.6567 [hep-th]} \BibitemShut
  {NoStop}%
\bibitem [{\citenamefont {Geyer}\ \emph {et~al.}(2016)\citenamefont {Geyer},
  \citenamefont {Mason}, \citenamefont {Monteiro},\ and\ \citenamefont
  {Tourkine}}]{Geyer:2016wjx}%
  \BibitemOpen
  \bibfield  {author} {\bibinfo {author} {\bibfnamefont {Y.}~\bibnamefont
  {Geyer}}, \bibinfo {author} {\bibfnamefont {L.}~\bibnamefont {Mason}},
  \bibinfo {author} {\bibfnamefont {R.}~\bibnamefont {Monteiro}}, \ and\
  \bibinfo {author} {\bibfnamefont {P.}~\bibnamefont {Tourkine}},\ }\href
  {\doibase 10.1103/PhysRevD.94.125029} {\bibfield  {journal} {\bibinfo
  {journal} {Phys. Rev. D}\ }\textbf {\bibinfo {volume} {94}},\ \bibinfo
  {pages} {125029} (\bibinfo {year} {2016})},\ \Eprint
  {http://arxiv.org/abs/1607.08887} {arXiv:1607.08887 [hep-th]} \BibitemShut
  {NoStop}%
\bibitem [{\citenamefont {Geyer}\ and\ \citenamefont
  {Monteiro}(2018)}]{Geyer:2018xwu}%
  \BibitemOpen
  \bibfield  {author} {\bibinfo {author} {\bibfnamefont {Y.}~\bibnamefont
  {Geyer}}\ and\ \bibinfo {author} {\bibfnamefont {R.}~\bibnamefont
  {Monteiro}},\ }\href {\doibase 10.1007/JHEP11(2018)008} {\bibfield  {journal}
  {\bibinfo  {journal} {JHEP}\ }\textbf {\bibinfo {volume} {11}},\ \bibinfo
  {pages} {008} (\bibinfo {year} {2018})},\ \Eprint
  {http://arxiv.org/abs/1805.05344} {arXiv:1805.05344 [hep-th]} \BibitemShut
  {NoStop}%
\bibitem [{\citenamefont {Geyer}\ \emph {et~al.}(2019)\citenamefont {Geyer},
  \citenamefont {Monteiro},\ and\ \citenamefont
  {Stark-Much\~ao}}]{Geyer:2019hnn}%
  \BibitemOpen
  \bibfield  {author} {\bibinfo {author} {\bibfnamefont {Y.}~\bibnamefont
  {Geyer}}, \bibinfo {author} {\bibfnamefont {R.}~\bibnamefont {Monteiro}}, \
  and\ \bibinfo {author} {\bibfnamefont {R.}~\bibnamefont {Stark-Much\~ao}},\
  }\href {\doibase 10.1007/JHEP12(2019)049} {\bibfield  {journal} {\bibinfo
  {journal} {JHEP}\ }\textbf {\bibinfo {volume} {12}},\ \bibinfo {pages} {049}
  (\bibinfo {year} {2019})},\ \Eprint {http://arxiv.org/abs/1908.05221}
  {arXiv:1908.05221 [hep-th]} \BibitemShut {NoStop}%
\bibitem [{\citenamefont {D'Hoker}\ \emph {et~al.}(2020)\citenamefont
  {D'Hoker}, \citenamefont {Mafra}, \citenamefont {Pioline},\ and\
  \citenamefont {Schlotterer}}]{DHoker:2020prr}%
  \BibitemOpen
  \bibfield  {author} {\bibinfo {author} {\bibfnamefont {E.}~\bibnamefont
  {D'Hoker}}, \bibinfo {author} {\bibfnamefont {C.~R.}\ \bibnamefont {Mafra}},
  \bibinfo {author} {\bibfnamefont {B.}~\bibnamefont {Pioline}}, \ and\
  \bibinfo {author} {\bibfnamefont {O.}~\bibnamefont {Schlotterer}},\ }\href
  {\doibase 10.1007/JHEP08(2020)135} {\bibfield  {journal} {\bibinfo  {journal}
  {JHEP}\ }\textbf {\bibinfo {volume} {08}},\ \bibinfo {pages} {135} (\bibinfo
  {year} {2020})},\ \Eprint {http://arxiv.org/abs/2006.05270} {arXiv:2006.05270
  [hep-th]} \BibitemShut {NoStop}%
\bibitem [{\citenamefont {Geyer}\ \emph {et~al.}(2021)\citenamefont {Geyer},
  \citenamefont {Monteiro},\ and\ \citenamefont
  {Stark-Much\~ao}}]{Geyer:2021oox}%
  \BibitemOpen
  \bibfield  {author} {\bibinfo {author} {\bibfnamefont {Y.}~\bibnamefont
  {Geyer}}, \bibinfo {author} {\bibfnamefont {R.}~\bibnamefont {Monteiro}}, \
  and\ \bibinfo {author} {\bibfnamefont {R.}~\bibnamefont {Stark-Much\~ao}},\
  }\href {\doibase 10.1103/PhysRevLett.127.211603} {\bibfield  {journal}
  {\bibinfo  {journal} {Phys. Rev. Lett.}\ }\textbf {\bibinfo {volume} {127}},\
  \bibinfo {pages} {211603} (\bibinfo {year} {2021})},\ \Eprint
  {http://arxiv.org/abs/2106.03968} {arXiv:2106.03968 [hep-th]} \BibitemShut
  {NoStop}%
\end{thebibliography}%

\end{document}